\documentclass[journal,twoside]{IEEEtran}
\usepackage{amsmath,amsfonts}
\usepackage{amsmath,amssymb,bm,booktabs,multirow,url}
\usepackage{algorithmic}
\usepackage{algorithm}
\usepackage{array}
\usepackage[caption=false,font=footnotesize,labelfont=rm,textfont=rm]{subfig}
\usepackage{textcomp}
\usepackage{stfloats}
\usepackage{url}
\usepackage{verbatim}
\usepackage{graphicx}
\usepackage{cite}
\usepackage{colortbl}
\usepackage{hyperref}
\usepackage{xcolor}
\usepackage{utfsym}
\usepackage{threeparttable}
\usepackage[usestackEOL]{stackengine}
\begin{document}

\title{Generalized Gaussian Model for\\Learned Image Compression}

\author{Haotian Zhang, Li Li, \IEEEmembership{Member, IEEE,} and Dong Liu, \IEEEmembership{Senior Member, IEEE}
\thanks{Date of current version \today. This work was supported by the Natural Science Foundation of China under Grants 62036005 and 62021001. We acknowledge the support of GPU cluster built by MCC Lab of Information Science and Technology Institution, USTC.

The authors are with the MOE Key Laboratory of Brain-Inspired Intelligent Perception and Cognition, University of Science and Technology of China, Hefei 230093, China (e-mail: zhanghaotian@mail.ustc.edu.cn; lil1@ustc.edu.cn; dongeliu@ustc.edu.cn). (\textit{Corresponding author: Dong Liu})}}

\maketitle

\begin{abstract}
In learned image compression, probabilistic models play an essential role in characterizing the distribution of latent variables. The Gaussian model with mean and scale parameters has been widely used for its simplicity and effectiveness. Probabilistic models with more parameters, such as the Gaussian mixture models, can fit the distribution of latent variables more precisely, but the corresponding complexity is higher. To balance the compression performance and complexity, we extend the Gaussian model to the generalized Gaussian family for more flexible latent distribution modeling, introducing only one additional shape parameter $\beta$ than the Gaussian model. To enhance the performance of the generalized Gaussian model by alleviating the train-test mismatch, we propose improved training methods, including $\beta$-dependent lower bounds for scale parameters and gradient rectification. Our proposed generalized Gaussian model, coupled with the improved training methods, is demonstrated to outperform the Gaussian and Gaussian mixture models on a variety of learned image compression networks.
\end{abstract}


\section{Introduction}
Image compression is one of the most fundamental problems in image processing and information theory. Over the past four decades, many researchers have worked on developing and optimizing image compression codecs. Most image compression codecs follow the transform coding scheme \cite{goyal2001transform}, where images are transformed to a latent space for decorrelation, followed by quantization and entropy coding. In traditional image compression codecs, such as JPEG \cite{wallace1991jpeg}, JPEG2000 \cite{skodras2001jpeg}, BPG \footnote{BPG is an image coding format based on HEVC intra coding, see \url{http://bellard.org/bpg/}.}, and VVC \cite{bross2021overview}, different modules are artificially designed and separately optimized. 

In recent years, learned image compression methods have achieved superior performance by exploiting the advantages of deep neural networks. In contrast to traditional approaches, learned image compression \cite{balle2016end} optimizes different modules in an end-to-end manner. A typical learned image compression method involves an analysis transform, synthesis transform, quantizer, and entropy model. First, the image is transformed into a latent representation through the analysis transform, and then the latent is quantized for digital transmission. The discrete latent is then losslessly coded by the entropy model to further reduce its size. Finally, the synthesis transform reconstructs an image from the discrete latent. These modules are jointly optimized to minimize the rate-distortion cost during training. 

In learned image compression, the probabilistic model is part of the entropy model and is essential in characterizing the distribution of latent variables. 
The degree of matching between the probabilistic model and the actual distribution of latent variables significantly influences the bitrate of compressed images. A more precise probabilistic model can reduce the bitrate. 
Probabilistic models with more parameters can fit the distribution of latent variables more precisely, but the corresponding complexity will also be higher.
In \cite{balle2018variational}, a zero-mean Gaussian scale model is used. The scale parameters are estimated based on side information. In \cite{minnen2018joint}, the probabilistic model is extended to the Gaussian Model (GM) with mean and scale parameters. Some studies further propose mixture probabilistic models, such as the Gaussian Mixture Model (GMM) with 9 parameters \cite{cheng2020learned} for more accurate distribution modeling. These mixture models improve compression performance compared to the Gaussian model while introducing more parameters that need to be estimated and higher complexity. 

\begin{figure}[!t]
  \centering
    \includegraphics[width=0.9\linewidth]{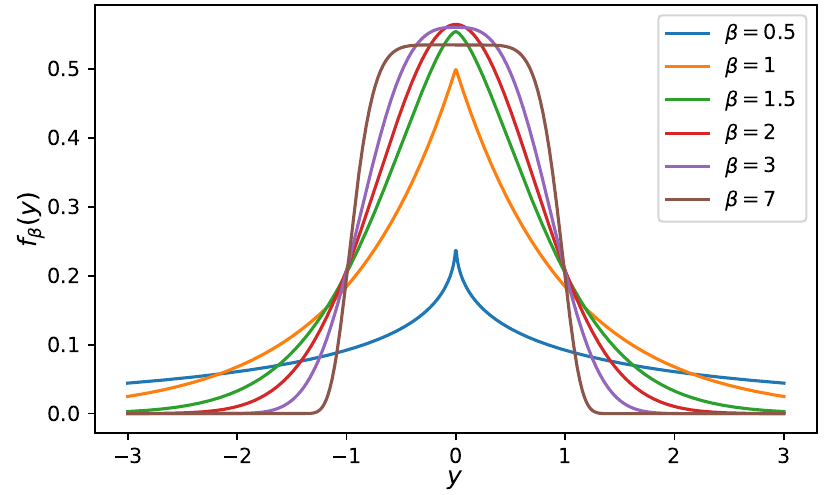}
    \caption{
    Shape of the Probability Density Function (PDF), as formulated by Eq. (\ref{eq:pdf_cdf}), of the Generalized Gaussian Model (GGM) with various shape parameters $\beta$. The mean and scale parameters are fixed as $\mu=0,\alpha=1$.
    }
    \label{fig:GGM_vary_beta}
\end{figure}

In this paper, to achieve a better balance between compression performance and complexity, we extend the Gaussian model with mean and scale parameters to the generalized Gaussian model. The Gaussian model is denoted as 
$Y\sim\mathcal{N}(\mu,\sigma^2)$,
where $\mu$ and $\sigma$ are the mean and scale parameters ($\sigma^2$ is the variance).
Compared to the Gaussian model, the Generalized Gaussian Model (GGM) offers a high degree of flexibility in distribution modeling by introducing only one additional shape parameter, which is denoted as
\begin{align}
    Y\sim\mathcal{N}_{\beta}(\mu,\alpha^{\beta}),
\end{align}
where $\mu$, $\alpha$, and $\beta$ are the mean, scale and shape parameters.
GGM degenerates into Gaussian when $\beta=2$ (with $\sigma=\alpha/\sqrt{2}$). As shown in Fig. \ref{fig:GGM_vary_beta}, GGM offers more flexible distribution modeling capabilities than Gaussian, particularly in handling data with varying degrees of tailing. 
We combine GGM with conditional entropy models by incorporating it into the end-to-end training process. 
To target different levels of complexity, we present three methods based on model-wise (GGM-m), channel-wise (GGM-c), and element-wise (GGM-e) shape parameters.

Enabling the end-to-end training requires incorporating non-differentiable quantization into the gradient-based training of the networks. Since the derivative of rounding is zero almost everywhere, preventing us from optimizing the analysis transform, the end-to-end training process for learned image compression usually replaces rounding by adding uniform noise \cite{balle2016end} for rate estimation. However, the discrepancy between noise relaxation during training and rounding during testing causes train-test mismatch, significantly affecting the compression performance. Zhang \textit{et al.} \cite{zhang2023uniform} show that with the Gaussian model applied, the train-test mismatch in the rate term is substantial for some small estimated scale parameters. Consequently, they suggest setting a proper lower bound for the estimated scale parameters to reduce the effect of train-test mismatch, resulting in better performance. We observe a similar phenomenon in GGM, while the degree of train-test mismatch varies across different $\beta$. To reduce the train-test mismatch when training with GGM, we propose improved training methods. First, we propose $\beta$-dependent lower bounds for scale parameters to adaptively mitigate the train-test mismatch across different regions. The bounds for scale parameters could effectively reduce the train-test mismatch, but it will also affect the optimization of shape parameters. To address this problem, we further propose a gradient rectification method to correct the optimization of shape parameters. In addition to the lower bound trick, Zhang \textit{et al.} \cite{zhang2023uniform} show that applying zero-center quantization \cite{minnen2020channel}, \textit{i.e.}, $\hat{y}=\lfloor y-\mu\rceil+\mu$, to the single Gaussian model can also reduce the negative influence of train-test mismatch. Our analyses show a similar property in GGM, so we also adopt zero-center quantization for GGM. With our improved training methods and the adoption of zero-center quantization, GGM could outperform GMM.

Experimental results demonstrate that our proposed GGM with improved training methods outperforms representative probabilistic models on various learned image compression methods. Our GGM-m outperforms GM with the same network complexity and comparable coding time. Our GGM-c achieves better performance with the same network complexity and longer coding time. Our GGM-e achieves further performance improvement with higher network complexity and longer coding time. The increase in coding time for GGM-c and GGM-e is less than 8\% compared to that of GM. With the help of zero-center quantization and look-up tables-based entropy coding, our GGM-e outperforms GMM with lower complexity.

Our main contributions can be summarized as follows:
\begin{itemize}
    \item We present the Generalized Gaussian Model (GGM) to characterize the distribution of latent variables, with only one additional parameter compared to the Gaussian model. We propose three methods based on model-wise (GGM-m), channel-wise (GGM-c), and element-wise (GGM-e) shape parameters, each targeting different levels of complexity.
    \item We propose improved training methods and adopt zero-center quantization to reduce the negative influence of train-test mismatch for GGM, which greatly enhances performance.
    \item Experimental results show that our GGM-m, GGM-c, and GGM-e outperform GM on various learned image compression models. Our GGM-e method outperforms GMM and has lower complexity.
\end{itemize}

\section{Related Work}
\subsection{Learned Image Compression}
Learned image compression has made significant progress and demonstrated impressive performance. In the early years, some studies focused on optimizing with separate consideration of distortion and rate \cite{toderici2015variable}. Balle \textit{et al.} \cite{balle2016end} formulated learned image compression as a joint rate-distortion optimization problem. The scheme in \cite{balle2016end} contains four modules: analysis transform, synthesis transform, quantizer, and entropy model. Most recent studies follow this joint rate-distortion optimization scheme for advanced compression performance. 

The architecture of the neural networks in transforms is essential. Recurrent neural networks are used in \cite{toderici2015variable,toderici2017full, lin2020spatial}. Balle \textit{et al.} \cite{balle2016end} proposed a convolution neural network-based image compression model. Chen \textit{et al.} \cite{chen2021end} introduced a non-local attention module to capture global receptive field. Cheng \textit{et al.} \cite{cheng2020learned} proposed a local attention module. 
Attention mechanism is also investigated in \cite{guo2021causal,zou2022devil}. Some studies \cite{zhu2022transformer,zou2022devil} tried to construct transformer-based transforms. Liu \textit{et al.} \cite{liu2023learned} combined transformer and convolution neural network. In addition to non-invertible transforms, several studies utilized invertible ones. Some studies \cite{ma2020end,dong2024wavelet} proposed trained wavelet-like transform. Xie \textit{et al.} \cite{xie2021enhanced} combined non-invertible and invertible networks. For practicality, some recent studies investigated \cite{wang2023evc,yang2023shallow,minnen2023advancing} much more lightweight transforms.

The entropy model is used to estimate the distribution of quantized latent, which contains two parts: probabilistic model and parameter estimation. The probabilistic model will be introduced in Sec. \ref{sec:probabilistic_model}. In this section, we introduce the parameter estimation module. A factorized prior is employed in \cite{balle2016end}. Balle \textit{et al.} \cite{balle2018variational} proposed a hyperprior model that parameterizes the distribution as a Gaussian model conditioned on side information. Minnen \textit{et al.} \cite{minnen2018joint} proposed a more accurate entropy model, jointly utilizing an autoregressive context model and hyperprior. Some studies \cite{lee2018context,chen2021end,guo2021causal,hu2021learning,jiang2023mlic} also focus on improving entropy models for advanced performance.
To balance compression performance and running speed, parallel checkerboard context model \cite{he2021checkerboard} and channel-wise autoregressive model \cite{minnen2020channel} are proposed. He \textit{et al.} \cite{he2022elic} combined checkerboard and unevenly grouped channel-wise context model. Some studies \cite{qian2022entroformer,mentzer2023m2t,li2023flexible} contributed to transformer-based context models.

Uniform scalar quantization is a widely adopted method in learned image compression, with rounding being used. Since the gradient of rounding is zero almost everywhere, the standard back-propagation is inapplicable during training. To enable end-to-end optimization, lots of quantization surrogates are proposed. Training with additive uniform noise \cite{balle2016end} is a popular approach for approximating rounding. In \cite{theis2017lossy}, straight-through estimator \cite{bengio2013estimating} is adopted for training, which applies stochastic rounding in the forward pass but uses a modified gradient in the backward pass. Minnen and Singh \cite{minnen2020channel} empirically proposed a mixed quantization surrogate, which uses noisy latent for rate estimation but uses rounded latent and straight-through estimator when passing a synthesis transform. This mixed surrogate outperforms adding uniform noise and has been widely adopted in recent studies.

\subsection{Probabilistic Models for Learned Image Compression}
\label{sec:probabilistic_model}
The probabilistic model plays an essential role in characterizing the distribution of latent variables. A more accurate probabilistic model can reduce the bitrate of compressed images.
In \cite{balle2016end}, the quantized latent variables are assumed to be independent and follow a non-parametric probabilistic model. Conditional entropy models are then introduced to improve the accuracy of the entropy model. In \cite{balle2018variational}, the authors proposed estimating the distribution of latent variables with a zero-mean Gaussian scale model, where a hyperprior module is used to estimate the scale parameter. In this scheme, the Cumulative Distribution Function (CDF) must be constructed dynamically during decoding. In \cite{minnen2018joint}, the probabilistic model is extended to the Gaussian model with mean and scale parameters. Cheng \textit{et al.} \cite{cheng2020learned} further proposed a more accurate Gaussian Mixture Model (GMM), the weighted average of multiple Gaussian models with different means and scales. Mentzer \textit{et al.} \cite{mentzer2018conditional} adopted Logistic mixture models. Fu \textit{et al.} \cite{fu2023learned} combined different types of distributions and proposed the Gaussian-Laplacian-Logistic Mixture Model (GLLMM).

For entropy coding, arithmetic coders are usually adopted to entropy code $\hat{y}$ into the bitstream. The encoder requires both $\hat{y}$ and its CDF as input, and the same CDF should be fed to the decoder to decompress $\hat{y}$ correctly. For conditional entropy models, the CDF of each element needs to be constructed dynamically during encoding and decoding, which has high computational cost, large memory consumption, and floating-point errors. 
The entropy coding can fail catastrophically if the CDF differs even slightly between the sender and receiver, which could be caused by the platform-dependent round-off errors in floating-point calculation. 
Round-off errors emerge in two ways. One is the inference of the parameter estimation module, and another is the calculation of the CDF. Balle \textit{et al.} \cite{balle2019integer} proposed to use integer networks to obtain discrete parameters and store the pre-computed CDF of each discrete parameter into look-up tables (LUTs) to avoid floating errors. In this way, the entropy coders only require the index, built upon the discrete parameter, of the corresponding CDF in LUTs, which has low computational cost, minimal memory consumption, and no floating-point errors. The study \cite{sun2021learned} extends the LUTs-based implementation from zero-mean Gaussian to mean-scale Gaussian with more LUTs. The study \cite{he2022post} tried to use LUTs-based implementation for GMM to avoid the floating point error. They followed \cite{sun2021learned} to share the LUTs across different Gaussian components. However, this approach also requires dynamically computing the CDF of each latent variable with the quantized weights. The number of CDF tables generated equals the number of latent variables, consuming higher memory cost and memory access time.

\subsection{Train-Test Mismatch in Learned Image Compression}
The train-test mismatch is caused by the discrepancy between noise relaxation during training and rounding during testing. 
The mixed surrogate \cite{minnen2020channel} uses noisy latent for rate estimation but uses rounded latent and straight-through estimator when passing a synthesis transform. Although the train-test mismatch in the distortion term is eliminated, the mismatch in the rate term exists. Some methods have also tried to use rounded latent and the straight-through estimator for rate estimation, resulting in poor performance \cite{guo2021soft, tsubota2023comprehensive} due to gradient bias. Zhang \textit{et al.} \cite{zhang2023uniform} showed that when the Gaussian model is applied, the mismatch in the rate term is substantial for some small estimated scale parameters. Consequently, they propose to set a proper lower bound for the estimated scale parameters during training to reduce the effect of train-test mismatch when optimizing the analysis transform, resulting in better compression performance. In addition, they suggest that when using mixed quantization surrogate during training, adopting zero-center quantization, \textit{i.e.}, $\hat{y}=\lfloor y-\mu \rceil+\mu$, helps to reduce the influence of train-test mismatch, thus resulting in improved performance.
Some studies \cite{guo2021soft,agustsson2020universally} also focus on eliminating the effect of train-test mismatch.

\section{Method}
\subsection{Characteristics of Generalized Gaussian Model}
\label{sec:formu_ggm}
The generalized Gaussian model encompasses probabilistic models such as the Gaussian and Laplacian models.
The Gaussian distribution is denoted as 
$Y\sim\mathcal{N}(\mu,\sigma^2)$,
where $\mu$ and $\sigma$ are the mean and scale parameters ($\sigma^2$ is the variance). Compared to the Gaussian model, the Generalized Gaussian Model (GGM) adds a shape parameter, which is denoted as $Y\sim\mathcal{N}_{\beta}(\mu,\alpha^{\beta})$,
where $\mu$, $\alpha$, and $\beta$ are the mean, scale and shape parameters. The Probability Density Function (PDF) is $\frac{\beta}{2\alpha\Gamma(1/\beta)}e^{-(\frac{|y-\mu|}{\alpha})^{\beta}}$. The standard PDF and Cumulative Distribution Function (CDF) (with $\mu=0$, $\alpha=1$) are formulated as
\begin{equation}\label{eq:pdf_cdf}
\begin{aligned}
     &f_{\beta}(y) = \frac{\beta}{2\Gamma(1/\beta)}e^{-|y|^{\beta}},\\
     c_{\beta}(y) = &\int_{-\infty}^{y}f_{\beta}(v) dv=\frac{1}{2}+\frac{\text{sgn}(y)}{2}P(\frac{1}{\beta},|y|^{\beta}),
\end{aligned}
\end{equation}
where $\Gamma(\cdot)$ denotes the gamma function
and $P(\cdot,\cdot)$ denotes the regularized lower incomplete gamma function\footnote{https://www.tensorflow.org/api\_docs/python/tf/math/igamma}. The GGM degenerates into Gaussian when $\beta=2$ (with $\sigma=\alpha/\sqrt{2}$) and Laplacian when $\beta=1$. The PDFs of GGM with various $\beta$ are shown in Fig. \ref{fig:GGM_vary_beta}. 
In GGM, the shape parameter $\beta$ controls the peakedness and tails of the PDF.

\begin{figure}[!t]
    \centering
    \includegraphics[width=\linewidth]{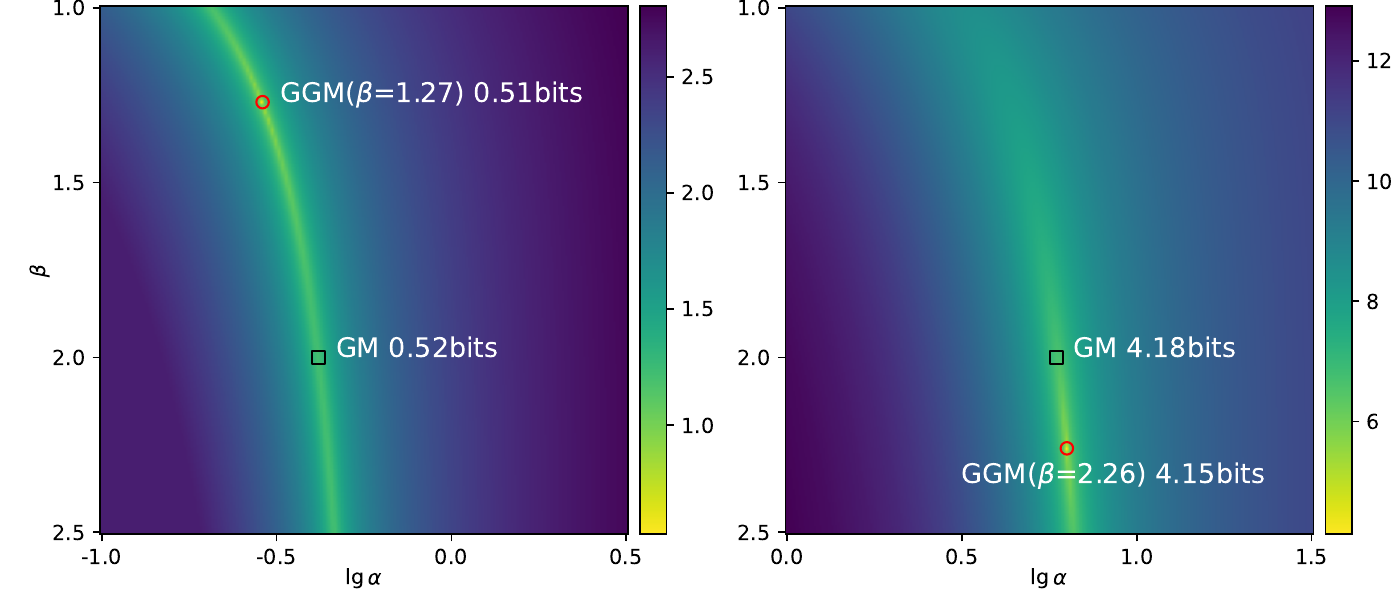}
    \caption{
    Bits estimated by GM and GGM of latent variables in mean-scale hyperprior model trained with GM. We train the mean-scale hyperprior model with Gaussian model \cite{minnen2018joint} and then collect latent variables (with the mean subtracted) with similar estimated scale parameter (Left: 66491 samples with $\alpha\in [0.42,0.425]$; Right: 12899 samples with $\alpha\in [5.6,5.7]$) from the Tecnick dataset. Then, we calculate the average bits of these latent variables under GGM with various $\beta$ and $\alpha$ parameters. The original parameters estimated by Gaussian are marked as $\square$, and the optimal parameters estimated by GGM are marked as $\color{red}\circ$. The visualization shows that even if the analysis transform is constrained by the Gaussian model, the actual distribution can also be better estimated by GGM.
    }
    \label{fig:fit_latent_distribution}
\end{figure}

\begin{figure}[!t]
    \centering
    \includegraphics[width=\linewidth]{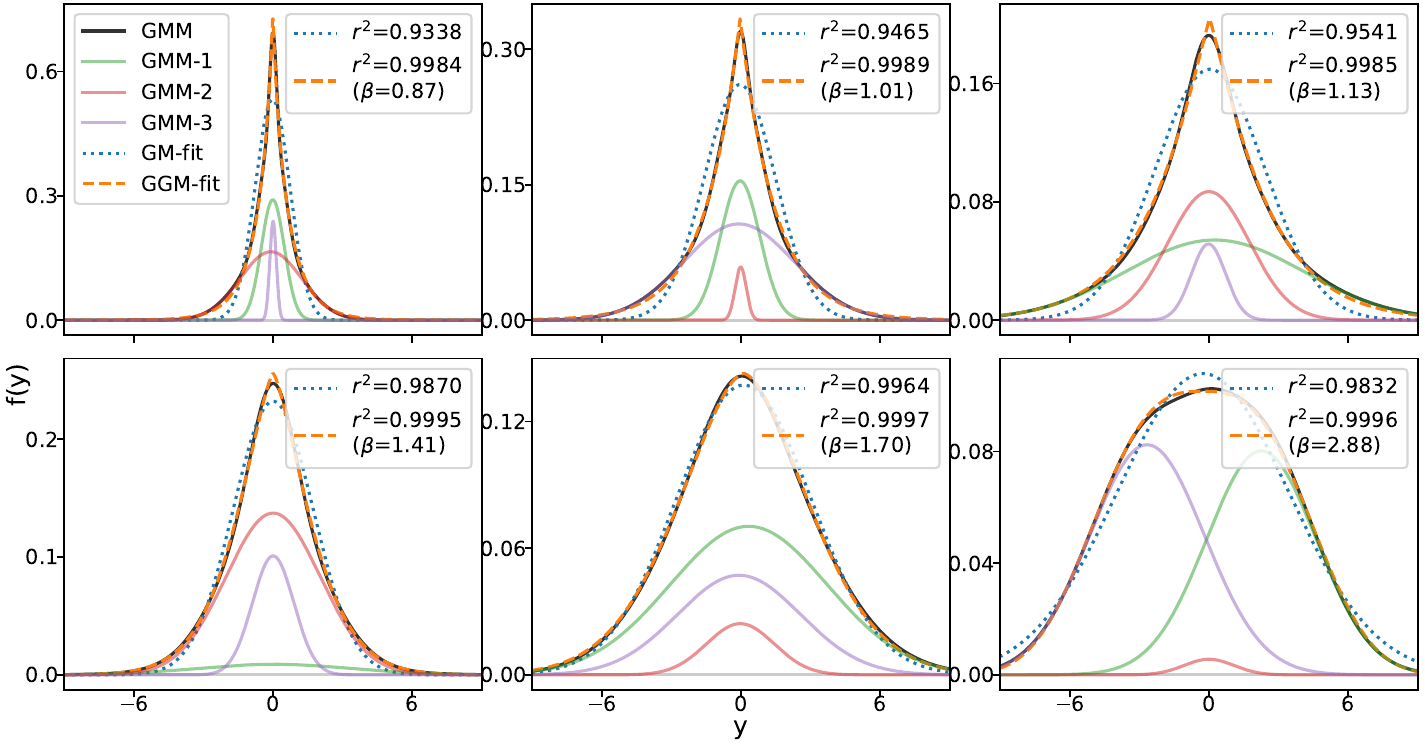}
    \caption{
    Visualization of the typical PDF of the latent variables estimated by the Gaussian Mixture Model (GMM) and the corresponding fitting with GGM and Gaussian model (GM). The results are collected from the hyperprior model with GMM as the probabilistic model. The plot of each Gaussian component in GMM (with label GGM-$i$, $i=1,2,3.$) is weighted. $r^2$ is a statistical measure used to assess the goodness of curve fitting, and the values range from 0 to 1, where 1 indicates a perfect fit and values closer to 1 indicate better fitting. The $r^2$ values fitted by GM and GGM (with its $\beta$ parameter) are shown on the upper right of each subplot. The visualizations show that GGM could fit the distributions of latent variables estimated by GMM well.
    }
    \label{fig:gmm}
\end{figure}

GGM offers more flexible distribution modeling capabilities than Gaussian, particularly in handling data with varying degrees of tailing. This generalized Gaussian family allows for tails that are either heavier than Gaussian (when $\beta<2$) or lighter than Gaussian (when $\beta >2$). As shown in Fig. \ref{fig:fit_latent_distribution}, even if the analysis transform is trained with the Gaussian model, the actual distribution of latent variables can also be better estimated by GGM, and a lower bitrate could be achieved. Additionally, GGM can  well fit the typical distribution of latent variables estimated by GMM, which approximate unimodal symmetric distributions, as shown in Fig. \ref{fig:gmm}. The goodness-of-fit of GGM notably surpasses that of the Gaussian model. 

\subsection{Generalized Gaussian Probabilistic Model for Learned Image Compression}
In the scheme of learned image compression \cite{balle2016end}, the sender applies an analysis transform to an image $x$, generating latent representation $y=g_a(x|\phi)$. Rounding is then applied to obtain the discrete latent $\hat{y}=Q(y)$. The discrete latent can be losslessly coded under an entropy model $q_{\hat{Y}}(\hat{y}|\psi)$. A non-adaptive factorized prior is typically used and shared between the encoder and decoder. Finally, the receiver recovers $\hat{y}$, and generates reconstruction $\hat{x}=g_s(\hat{y}|\theta)$ through the synthesis transform. The notations $\phi$, $\theta$, and $\psi$ are trainable parameters of the analysis transform, synthesis transform, and entropy model, respectively. In this paper, we use uppercase letters, like $Y$, to denote random variables and use lowercase letters, like $y$, to denote a sample of random variables without distinguishing between scalars and vectors.

In the study \cite{balle2018variational}, a hyperprior entropy model is proposed by introducing side information $z$ to capture redundancy in the latent representations $y$. The process can be formulated as 
\begin{align}
z = h_a(y|\phi_h),~\hat{z} = Q(z),~q_{\hat{Y}|\hat{Z}}(\hat{y}|\hat{z}) \leftarrow h_s(\hat{z}|\theta_h),
\end{align}
where $h_a$ and $h_s$ are the analysis and synthesis transform in the hyperprior auto-encoder, $\phi_h$ and $\theta_h$ are trainable parameters. $q_{\hat{Y}|\hat{Z}}$ is the estimated distribution of $\hat{Y}$ conditioned on the side information $\hat{Z}$. Following that, the study \cite{minnen2018joint} proposed a more accurate entropy model, which jointly utilizes an autoregressive context model and hyperprior module. In this conditional entropy coding scheme, the PDF of $\hat{y}$ is calculated dynamically during decoding, which introduces extra computational complexity. Therefore, a simple but effective probabilistic model used to characterize the distribution of $\hat{y}$ is crucial for conditional entropy models. 

Lots of efforts have been made towards more effective probabilistic models. 
The Gaussian scale model is used in \cite{balle2018variational},
\begin{align}
    q_{\hat{Y}|\hat{Z}}(\hat{y}|\hat{z})\sim \mathcal{N}(0,\sigma^2),
\end{align}
where $\sigma=h_s(\hat{z}|\theta_h)$. The study \cite{minnen2018joint} extends it to the Gaussian model with mean and scale parameters, 
\begin{align}
    q_{\hat{Y}|\hat{Z}}(\hat{y}|\hat{z})\sim \mathcal{N}(\mu,\sigma^2),
\end{align}
where $\{\mu,\sigma\}=h_s(\hat{z}|\theta_h)$,
which is widely used in recent studies. To further improve the performance, the Gaussian mixture model (GMM) is adopted in \cite{cheng2020learned}. The weighted average of multiple Gaussian models with different means and scales is used to estimate the distribution of $\hat{y}$,
\begin{align}
    q_{\hat{Y}|\hat{Z}}(\hat{y}|\hat{z})\sim\sum_{k=1}^{K} \omega_k \mathcal{N}(\mu_k,\sigma_k^2),
\end{align}
where $\omega_k$ is the weight of different Gaussian components and $\{\mu_k,\sigma_k,\omega_k(k=1,2,\cdots,K)\}=h_s(\hat{z}|\theta_h)$. Moreover, different types of distributions are combined in the Gaussian-Laplacian-Logistic Mixture Model (GLLMM) \cite{fu2023learned}. 

\begin{figure*}[!t]
    \centering
    \includegraphics[width=0.99\linewidth]{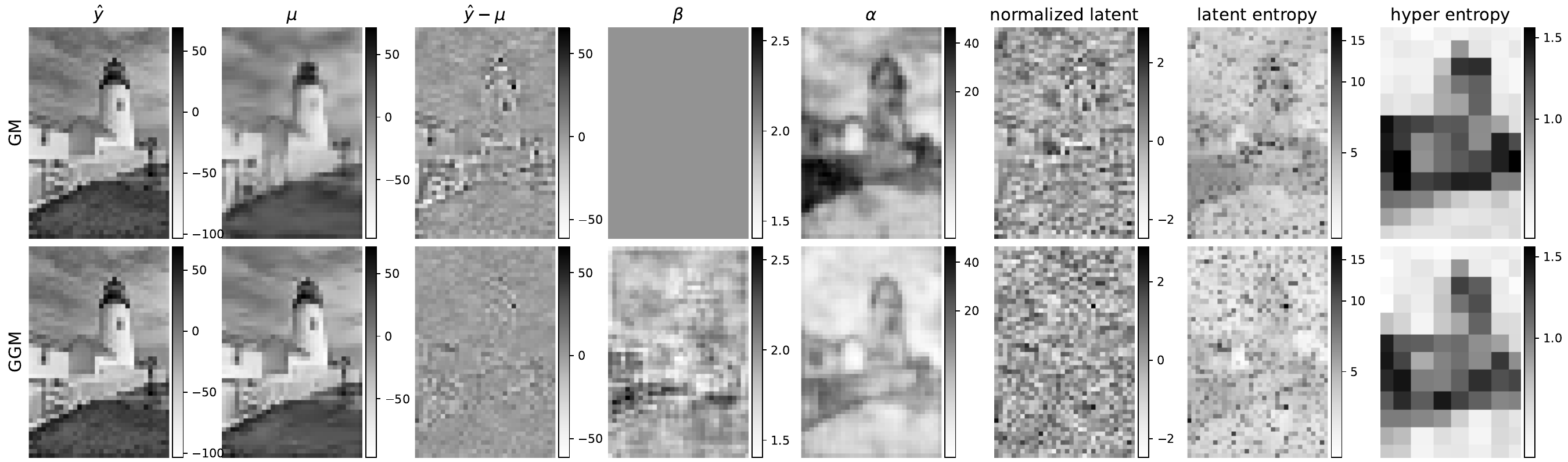}
    \caption{Visualization of the information for the channel with the highest entropy trained with GM and GGM (element-wise shape parameters) using image kodim19 in the Kodak dataset. 
    The bitrates of hyperprior, \textit{i.e.}, summation of \textit{hyper entropy}, are GM: 0.028bpp, GGM: 0.027bpp.
    The visualizations show that compared to GM, GGM reduces the prediction error, requires smaller scale parameters, and removes more structure from the normalized latent with the smaller hyperprior bitrate, which directly translates to a lower bitrate. The backbone model is the mean-scale hyperprior model \cite{minnen2018joint}. The normalized latent variables are first converted to uniform and then converted to Gaussian for visualization, $\hat{y}_{\text{norm}}=c_{\beta=2}^{-1}(c_{\beta}(\frac{\hat{y}-\mu}{\alpha}))$, where $c_{\beta}$ is the CDF of GGM as formulated in Eq. (\ref{eq:pdf_cdf}), and $c^{-1}_{\beta}$ is the inverse function of $c_{\beta}$.
    }
    \label{fig:19_visual}
\end{figure*}

These mixture models are more effective than the single Gaussian model because they introduce more parameters that need to be estimated by the conditional entropy model. However, these additional parameters also increase complexity. 
To better balance the compression performance and complexity, we extend the Gaussian model to the generalized Gaussian model with only one additional parameter for learned image compression,
\begin{align}
    q_{\hat{Y}|\hat{Z}}(\hat{y}|\hat{z})\sim \mathcal{N}_{\beta}(\mu,\alpha^{\beta}).
\end{align}
The rate of $\hat{y}$ is calculated through 
\begin{align}
    &R(\hat{y}) = R(\lfloor y\rceil) = -\log_2{q_{\hat{Y}|\hat{Z}}(\hat{y}|\hat{z})}\\
    &q_{\hat{Y}|\hat{Z}}(\hat{y}|\hat{z})=c_{\beta}(\frac{\hat{y}-\mu+0.5}{\alpha})-c_{\beta}(\frac{\hat{y}-\mu-0.5}{\alpha}),
\end{align}
where $c_{\beta}$ is the CDF of GGM, as depicted in Sec. \ref{sec:formu_ggm}. 
When optimizing the parameters $\mu, \alpha, \beta$ through gradient descent, the derivative of $c_{\beta}(y)$  can be calculated through
\begin{equation}
\begin{aligned}\label{eq:grad_cbeta}
&\frac{\partial c_{\beta}(y)}{\partial y} = \frac{\beta}{2\Gamma(1/\beta)}e^{-|y|^{\beta}},\\
&\frac{\partial c_{\beta}(y)}{\partial \beta} = \frac{\text{sgn}(y)}{2}(-\frac{1}{\beta^2}P^{'}+\frac{|y|\ln|y|}{\Gamma(1/\beta)}e^{-|y|^{\beta}}).
\end{aligned}
\end{equation}
For $P^{'}=\partial (P(\frac{1}{\beta},|y|^{\beta}))/\partial (1/\beta)$, we follow the implementation in Tensorflow. 
The derivation of Eq. (\ref{eq:grad_cbeta}) is included in Appendix \ref{supp:1}.

There are several ways to incorporate GGM with the conditional entropy model by varying the diversity of $\beta$. In this paper, we present three methods, each targeting different levels of complexity:
\subsubsection{Model-wise $\beta$ (\textbf{GGM-m})} For simplicity, we can set one global shape parameter for a model. All elements with different means and scales share the same $\beta_0$. After training, the $\beta_0$ is fixed. For GGM-m, only the mean and scale parameters are generated from the conditional entropy model, which does not introduce additional network complexity.
\begin{align}
    &q_{\hat{Y}|\hat{Z}}(\hat{y}_{kij}|\hat{z})\sim \mathcal{N}_{\beta_0}(\mu_{kij},\alpha_{kij}^{\beta_0}),\\
    &\{\mu,\alpha\}=h_s(\hat{z}|\theta_h),
\end{align}
where $k$ denotes the channel index of latent variables, and $\{i,j\}$ denote the spatial position.
\subsubsection{Channel-wise $\beta$ (\textbf{GGM-c})} The channel energy distribution exhibits distinct characteristics. To set one shape parameter for each channel could adjust the estimated distribution based on the characteristics of each channel. The elements inside one channel share the same $\beta_k$. After training, the $\beta_k$ of each channel is fixed. For GGM-c, only the mean and scale parameters are generated from the conditional entropy model, which does not introduce additional network complexity.
\begin{align}
    &q_{\hat{Y}|\hat{Z}}(\hat{y}_{kij}|\hat{z})\sim \mathcal{N}_{\beta_k}(\mu_{kij},\alpha_{kij}^{\beta_k}),\\
        &\{\mu,\alpha\}=h_s(\hat{z}|\theta_h),
\end{align}
\subsubsection{Element-wise $\beta$ (\textbf{GGM-e})}
For more flexible modeling, each element in latent representations could have one shape parameter. For GGM-e, the output dimension of the entropy parameter module is increased, resulting in increased network complexity. Each element's $\beta_{kij}$ is adaptive based on the conditional entropy model.
\begin{align}
    &q_{\hat{Y}|\hat{Z}}(\hat{y}_{kij}|\hat{z})\sim \mathcal{N}_{\beta_{kij}}(\mu_{kij},\alpha_{kij}^{\beta_{kij}}),\\
    &\{\mu,\alpha,\beta\}=h_s(\hat{z}|\theta_h).
\end{align}
The visualization of our GGM-e approach is shown in Fig. \ref{fig:19_visual}. Compared to Gaussian, GGM reduces the prediction error, requires smaller scale parameters, and removes more structure from the normalized latent with a smaller hyperprior bitrate, which translates to a lower bitrate.

\subsection{Improved Training Methods}
Enabling the end-to-end training requires incorporating non-differentiable quantization into the gradient-based training of the networks. 
Since the derivative of rounding is zero almost everywhere, preventing us from optimizing the analysis transform, the end-to-end training in learned image compression usually replaces rounding by adding uniform noise \cite{balle2016end, minnen2020channel} for rate estimation.
However, the discrepancy between noise relaxation during training and rounding during testing causes train-test mismatch, significantly affecting the compression performance.
Zhang \textit{et al.} \cite{zhang2023uniform} show that with the Gaussian model applied, the mismatch in the rate term is substantial for some small estimated scale parameters. Thus, they propose to set a proper lower bound for the scale parameters estimated by the entropy model to reduce the negative influence of train-test mismatch on optimizing the analysis transform, which results in better performance. 
GGM is an extension of the Gaussian model, and many of its properties are similar to those of the Gaussian model. By analyzing its properties, we observe a similar phenomenon in GGM, while the degree of train-test mismatch varies across different $\beta$. To reduce the train-test mismatch when training with GGM, we propose $\beta$-dependent lower bounds for scale parameters to adaptively mitigate the train-test mismatch across different regions. The bounds for scale parameters could effectively reduce the train-test mismatch, but it will also affect the optimization of shape parameters. To address this problem, we further propose a gradient rectification method to correct the optimization of shape parameters.

\subsubsection{$\beta$-dependent lower bound for scale parameter}
\label{sec:mismatch}
\begin{figure}[!t]
  \centering
  \subfloat[Visualization of rate estimation error $(R(\tilde{Y})-R(\lfloor Y\rceil))/R(\lfloor Y\rceil)$.]
    {  
        \includegraphics[width=0.94\linewidth]{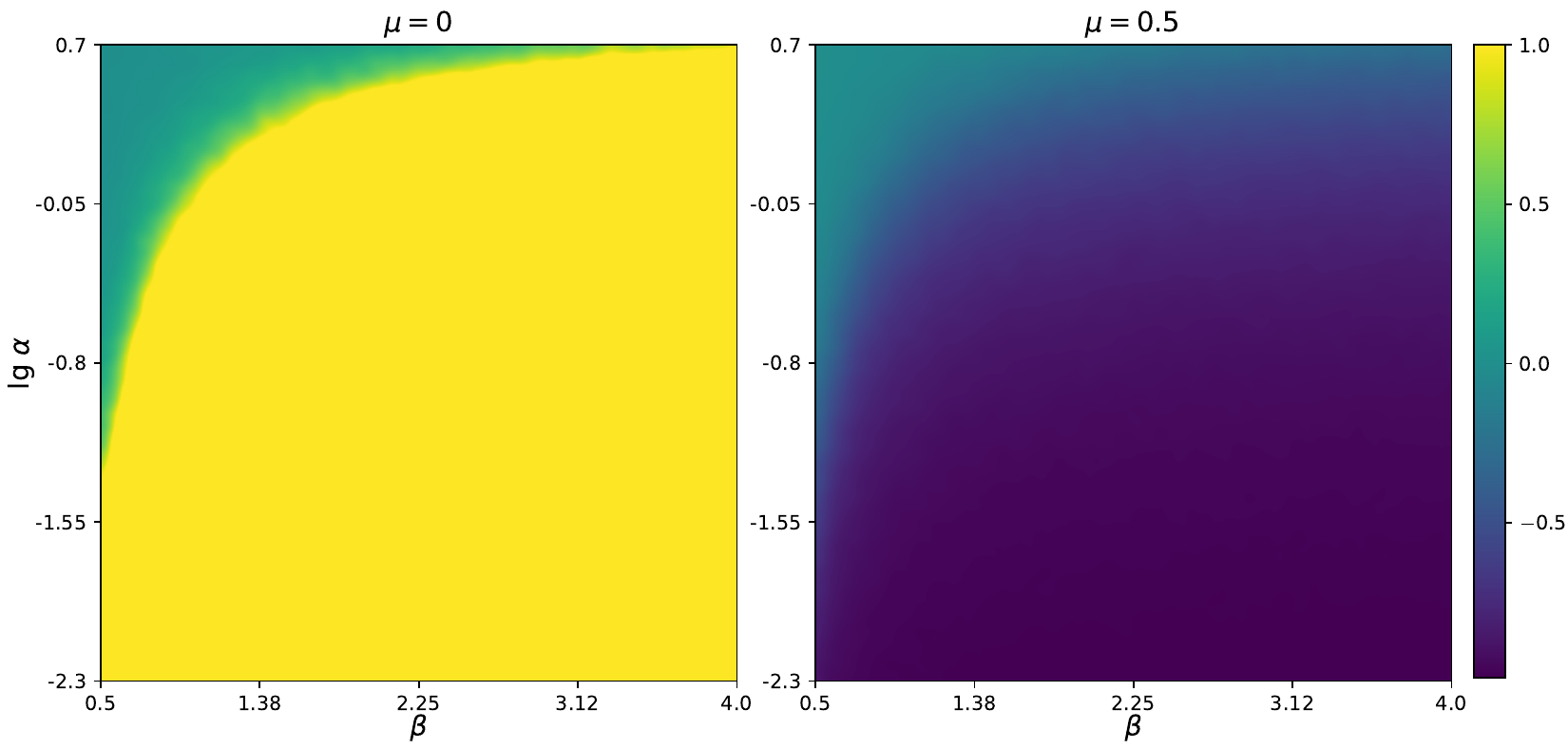}
        \label{fig:rate_estimation}
    }

    \subfloat[Distribution of learned $\beta-\alpha$.]
    {  
        \includegraphics[width=0.475\linewidth]{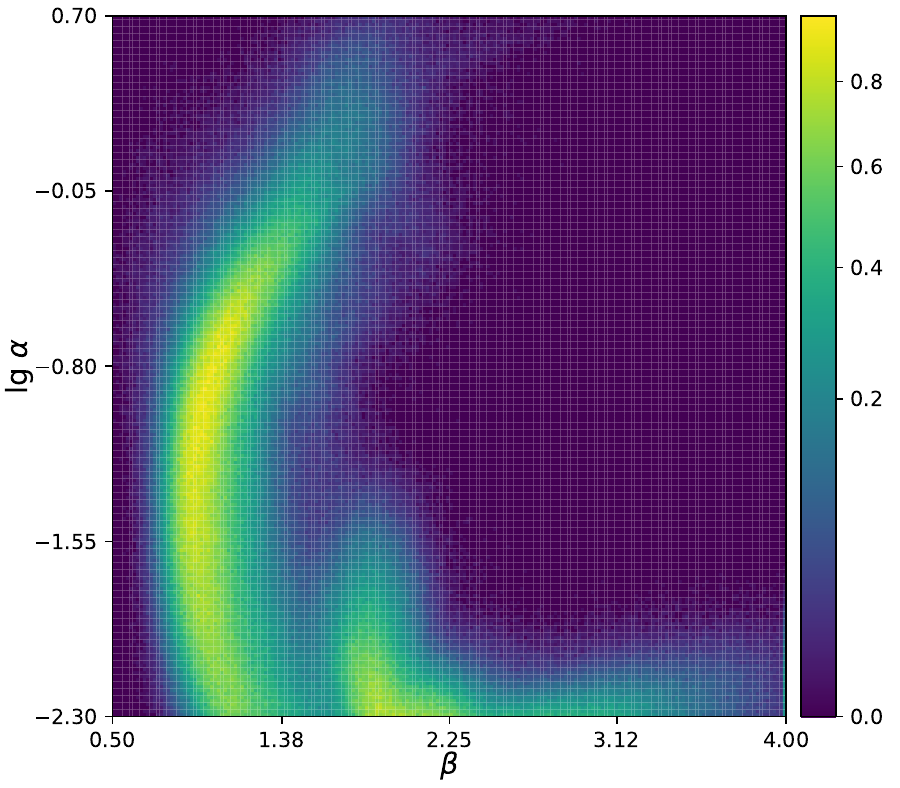}
        \label{fig:distribution_scale}
    } 
    \hspace{-0.5em}
    \subfloat[Scale lower bound for GGM.]
    {  
        \includegraphics[width=0.475\linewidth]{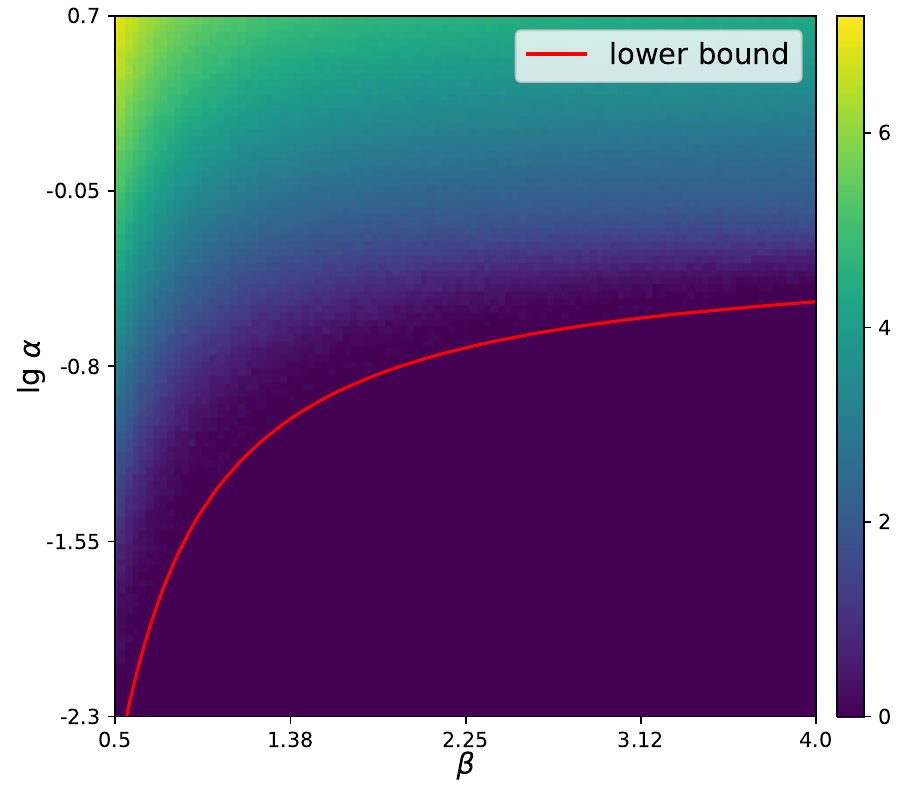}
        \label{fig:lower_bound}
    } 
    
    \caption{
   Illustration of the $\beta$-dependent lower bound for scale parameter. (a) shows the visualization of $\Delta R=(R(\tilde{Y})-R(\lfloor Y\rceil))/R(\lfloor Y\rceil)$ with various GGM distributions, as formulated by Eq. (\ref{eq:rate_estimation}). Values of $\Delta R$ greater than 1 in (a) are clipped. (b) shows the distribution of shape and scale parameters trained with GGM-e on the mean-scale hyperprior model \cite{minnen2018joint}. The distribution is collected from the Kodak dataset with 7077888 samples. (c) shows the visualization of $\beta$-dependent lower bound for scale parameter and the corresponding rate with $\mu=0$ estimated with rounding.}
    \label{fig:mse}
\end{figure}

We first analyze the train-test mismatch in the rate term for GGM. For simplicity, we assume that the estimated distribution perfectly matches the actual distribution.
Figure \ref{fig:rate_estimation} shows the distinction between the average rate of $Y$ 
 estimated by noisy relaxation $R(\tilde{Y})$ and rounding $R(\lfloor Y\rceil)$,
\begin{equation}\label{eq:rate_estimation}
 \begin{aligned}
     Y\sim\mathcal{N}_{\beta}&(\mu,\alpha^{\beta}),~\tilde{Y} = Y+U,~U\sim\mathcal{U}(-0.5,0.5),\\
     &R(\tilde{Y}) = \mathbb{E}[-\log_2{q_{\tilde{Y}|\hat{Z}}(\tilde{y}|\hat{z})}],\\
     R&(\lfloor Y\rceil) = \mathbb{E}[-\log_2{q_{\hat{Y}|\hat{Z}}(\hat{y}|\hat{z})}].\\
 \end{aligned}
\end{equation}
We notice that the mismatch in the rate term is substantial for some small estimated scale parameters. For instance, when $\mu=0$, most estimated rates with small scale parameters are larger than the actual rate, while for $\mu=0.5$, most estimated rates with small scale parameters are much smaller than the actual rate. Figure \ref{fig:distribution_scale} further shows that the scale parameters of the GGM trained to model the distribution of latent variables are concentrated on some small values where the mismatch is enormous. Therefore, similar to the Gaussian model, it is also necessary to set a proper lower bound for the scale parameter of GGM to reduce the influence of train-test mismatch. Additionally, the scale range resulting in substantial mismatch varies across different $\beta$ as shown in Fig. \ref{fig:rate_estimation}. Consequently, the optimal lower bound for the scale parameter should vary across $\beta$.

We denote the bounded scale parameter as 
\begin{equation}\label{eq:bound}
\begin{aligned}
\alpha_{b}=
\begin{cases}
\alpha, \text{if } \alpha>\alpha_{\beta};\\
\alpha_{\beta}, \text{if } \alpha\leq \alpha_{\beta},
\end{cases}
\end{aligned}
\end{equation}
where 
$\alpha_{\beta}$ is the $\beta$-dependent lower bound.
To determine the optimal $\alpha_\beta$ for each $\beta$ through experiments is too expensive. The results in \cite{zhang2023uniform} suggest that superior performance can be achieved by setting the bound around the value, for scale parameters smaller than which the quantized CDF remains unchanged. A larger bound risks excluding the actual distribution of $\hat{y}$ from the feasible estimated distributions, while a smaller one provides inadequate mitigation of train-test mismatch.
Therefore, we determine the proper bounds $\alpha_\beta$ through 
\begin{equation}\label{eq:scale_bound}
\begin{aligned}
\alpha_{\beta} = \max_{\alpha}\{\alpha: c_{\beta}(\frac{0.5}{\alpha})-c_{\beta}(\frac{-0.5}{\alpha})>1-10^{-5}\}.
\end{aligned}
\end{equation}
The result is shown in Fig. \ref{fig:lower_bound}, where $\alpha_\beta$ increases as $\beta$ increases. 

Moreover, since the 
$\beta$-dependent lower bound is used during training, where $\beta$ changes continuously, dynamically determining $\alpha_\beta$ through searching based on Eq. (\ref{eq:scale_bound}) is time-consuming. To address this problem, we use one small network to fit $\alpha_\beta$ and then freeze it when training learned image compression models. The detailed architecture of this module is included in Appendix \ref{supp:3}. It's important to note that this module is only utilized during training without slowing the testing speed.

\subsubsection{Gradient rectification}
\label{sec:grad}

\begin{figure*}[!t]
  \centering
    \includegraphics[width=0.99\linewidth]{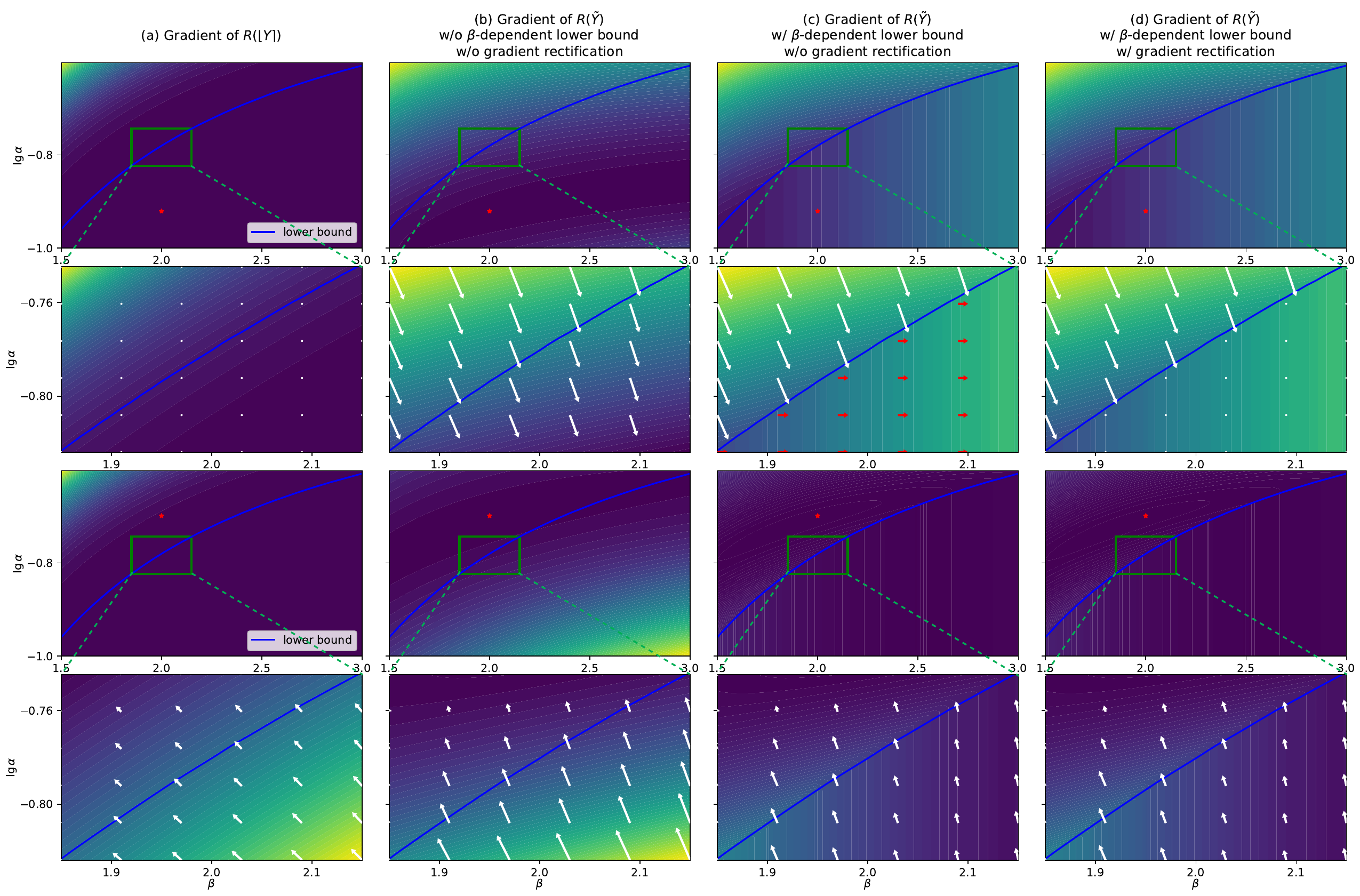}
    \caption{
   Visualization of the gradient of rate to $\alpha$ and $\beta$ with different methods. The direction of the arrow is the negative direction of the gradient, \textit{i.e.}, direction to minimize rate. The wrong direction is noted as \textcolor{red}{red} arrows. The length of the arrow represents the magnitude of the gradient. The gradient direction in (d) is the corrected gradient during training. In the first two rows, $Y\sim\mathcal{N}_{2}(0,0.12^2)$. In the last two rows, $Y\sim\mathcal{N}_{2}(0,0.2^2)$. The actual distribution location of $Y$ is marked as $\color{red}\star$. The mean parameter is assumed to be zero. The second and fourth rows are zoomed-in views of the first and third rows.}
\label{fig:grad_rectify}
\end{figure*}

Using $\beta$-dependent lower bound for the scale parameter can reduce the train-test mismatch, while it affects the optimization of shape parameters. We present an example to show how it hurts the optimization of shape parameters. We consider the scenario where the latent variable's distribution lies below the bound. As shown in Fig. \ref{fig:grad_rectify}a, if the actual distribution lies below the lower bound, the gradients of $R(\lfloor Y \rceil)$ to the estimated $\alpha$ and $\beta$, which lie below the lower bounds, are almost zero since their estimated rate is nearly the same. For $R(\tilde{Y})$, the gradient always leads $\alpha$ and $\beta$ towards the actual distribution by minimizing the rate, as shown in Fig. \ref{fig:grad_rectify}b. 
After using the proposed $\beta$-dependent lower bound, since the scale parameter can not further go smaller, the gradient to the scale parameter is eliminated as formulated in Eq. (\ref{eq:grad_bound}). 
\begin{equation}\label{eq:grad_bound}
\begin{aligned}
    & \text{For } (\alpha,\beta) \in \{\alpha,\beta:\alpha<\alpha_{\beta}\},\quad \alpha_{b}=\alpha_{\beta}, \\
    &\frac{\partial R(\tilde{Y};\alpha_{b},\beta)}{\partial \alpha} =
\begin{cases}
\eta, \text{if } \eta\leq 0;\\
0, \text{if }\eta > 0,
\end{cases}\\
&\text{where }\eta=\frac{\partial R(\tilde{Y};\alpha_{b},\beta)}{\partial \alpha_{b}}|_{\alpha_b=\alpha_{\beta}},\\
    &\frac{\partial R(\tilde{Y};\alpha_{b},\beta)}{\partial \beta} = \frac{\partial R(\tilde{Y};\alpha_{\beta},\beta)}{\partial \beta},
\end{aligned}
\end{equation}
Only the gradient to the shape parameter is preserved, and the same with the gradient at the bounded value. However, this gradient tends to make the shape parameter much larger, as shown in Fig. \ref{fig:grad_rectify}c. This is because the same scale parameter with a larger shape parameter results in smaller entropy.
This wrong direction gradient will hurt the optimization of the parameters of the probabilistic model, resulting in larger shape parameters than needed. 

To address this problem, we propose a joint gradient rectification method for optimizing shape and scale parameters by only preserving gradients that lead above the bound when the parameters already lie below the curve. Based on Eq. (\ref{eq:grad_bound}), we rectify the gradient to $\beta$ through
\begin{equation}\label{eq:grad_beta}
\begin{aligned}
    &\frac{\partial R(\tilde{Y};\alpha_{b},\beta)}{\partial \beta} =
\begin{cases}
0, \text{if } \zeta \leq 0;\\
\zeta, \text{if }\zeta > 0,
\end{cases}\text{where }\zeta=\frac{\partial R(\tilde{Y};\alpha_{\beta},\beta)}{\partial \beta}.
\end{aligned}
\end{equation}
The gradient after rectification is shown in Fig. \ref{fig:grad_rectify}d. This method eliminates the wrong-direction gradient, and the right-direction gradient could be preserved if the latent variable's distribution lies above the bound. Introducing this gradient rectification method does not significantly slow the training process while improving compression performance. The training time and loss convergence curve are included in Appendix \ref{supp:4}.

\subsection{Implementation}
\subsubsection{Zero-center quantization}
We follow previous studies \cite{minnen2020channel,he2022elic,liu2023learned} to use the mixed quantization surrogate \cite{minnen2020channel} for training learned image compression models. Furthermore, following the suggestion in \cite{minnen2020channel,zhang2023uniform}, we adopt the zero-center quantization to GGM, \textit{i.e.}, $\hat{y}=\lfloor y-\mu\rceil+\mu$. For this method, the symbol $\lfloor y-\mu\rceil$ is coded into the bitstream using the CDF constructed based on the corresponding $\beta$ and $\alpha$. The decoder reconstructs $\hat{y}$ as $\lfloor y-\mu\rceil+\mu$. 

When training with the mixed quantization surrogate, which uses noisy latent for rate estimation but uses the rounded latent to calculate distortion, there is only a mismatch in the rate term. With the assumption that the estimated distribution of latent variables perfectly matches the actual distribution, the distribution of $Y-\mu$ is a zero-mean GGM. As shown in Fig. \ref{fig:rate_estimation}, for zero-center quantization, \textit{i.e.}, coding zero-mean GGM variables $Y-\mu$, the rate estimated by noisy latent is larger than the actual rounded rate, providing an upper bound for the true rate-distortion cost. Optimizing this upper bound could lead to improvement in true compression performance. For nonzero-center quantization, \textit{i.e.}, coding nonzero-mean variables $Y$, the rate estimated by noisy latent may be significantly smaller than the actual rounded rate, as shown in Fig. \ref{fig:rate_estimation}. Thus, a decrease in the relaxed loss function may not guarantee a decrease in the true loss. Therefore, applying zero-center quantization for GGM could reduce the negative influence of train-test mismatch compared to the nonzero-center quantization, achieving better compression performance.
\subsubsection{LUTs-based implementation for entropy coding}
\label{sec:lut}
We follow previous studies \cite{balle2019integer,sun2021learned,begaint2020compressai} to adopt the look-up tables-based (LUTs-based) entropy coding for GM and extend this approach for GGM.
The LUTs-based method only influences the test stage and does not influence the training process. By using this approach, the cumulative distribution function (CDF) tables for entropy coding are pre-computed, eliminating the need for dynamic calculation during encoding and decoding, thereby saving computational costs and reducing coding time.

We present the overview of the LUTs-based entropy coding below. After training, several entropy parameter values of the probabilistic model are sampled in a specific manner from their respective possible ranges. We then calculate the corresponding CDF table for each sampled value. The encoder and decoder share these pre-computed CDF tables. During the actual encoding and decoding process, the predicted entropy parameters, which model the distribution of latent variables, are quantized to the nearest sampled value. These quantized values are then used to index the corresponding CDF table, which is necessary for entropy coders. For both GM and GGM, which have an identifiable mean parameter, we follow previous studies \cite{minnen2020channel,he2022elic,liu2023learned} to apply zero-center quantization, where the symbols encoded into bitstreams are given by $\lfloor y - \mu \rceil$. For GM, the entropy parameter involved in entropy coding is $\sigma$, while for GGM, the entropy parameters involved in entropy coding are $\beta$ and $\alpha$.

For GGM-m, since there is only one $\beta$ value, we only sample one $\beta$ value. In addition, we linearly sample 160 values in the log-scale field of $[0.01,60]$ for the $\alpha$ parameter. 
For GGM-c and GGM-e, the $\beta$ parameter is linearly sampled from the range $[0.5, 3]$ with 20 samples, and the $\alpha$ parameter is linearly sampled in the log-scale range $[0.01, 60]$ with 160 samples. We then combine the 20 $\beta$ values and 160 $\alpha$ values to generate 3200 $\beta-\alpha$ pairs and calculate the corresponding CDF table for each pair. The precision of the CDF table is set to 16-bit unsigned integer (uint16), with a maximum length of each CDF table set to 256. The experimental results for determining the number of CDF tables and more details about the LUTs-based entropy coding are included in Appendix \ref{supp:5}.

\begin{figure*}[!t]
  \centering
  \subfloat[All]
    {  
        \includegraphics[width=0.24\linewidth]{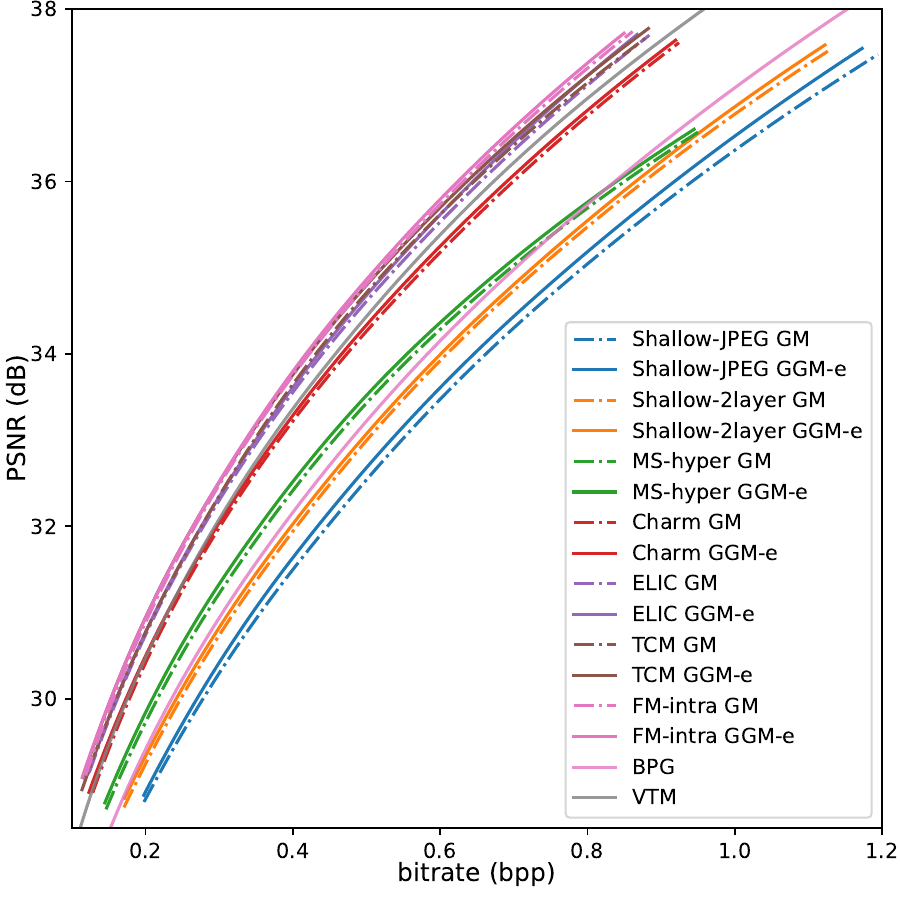}
        \label{fig:mse_single}
    } 
    \subfloat[Shallow-JPEG\cite{yang2023shallow}]
    {  
        \includegraphics[width=0.24\linewidth]{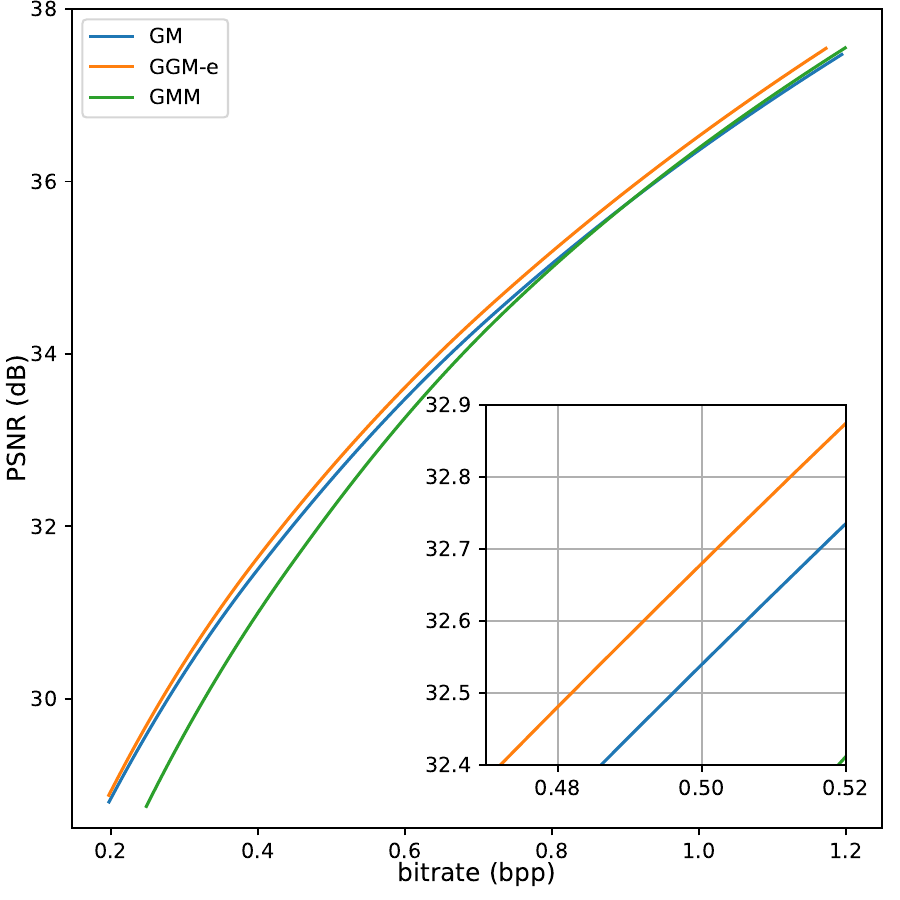}
        \label{fig:mse_single}
    } 
    \subfloat[Shallow-2layer\cite{yang2023shallow}]
    {  
        \includegraphics[width=0.24\linewidth]{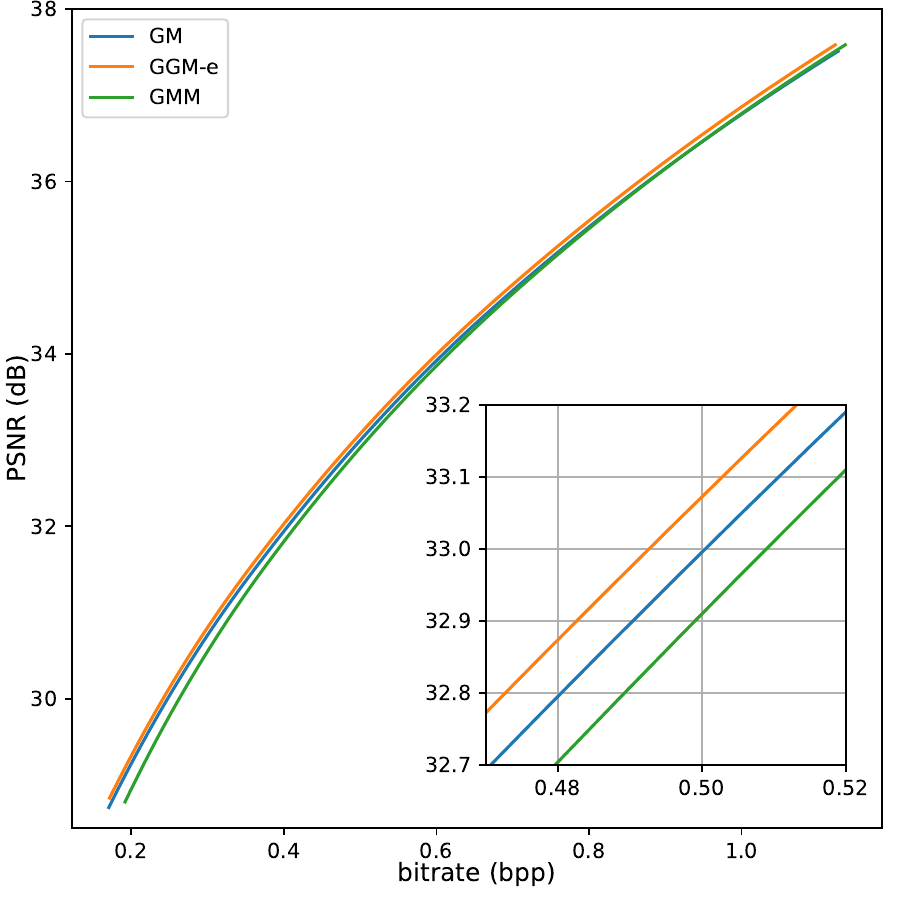}
        \label{fig:mse_single}
    } 
    \subfloat[MS-hyper\cite{minnen2018joint}]
    {  
        \includegraphics[width=0.24\linewidth]{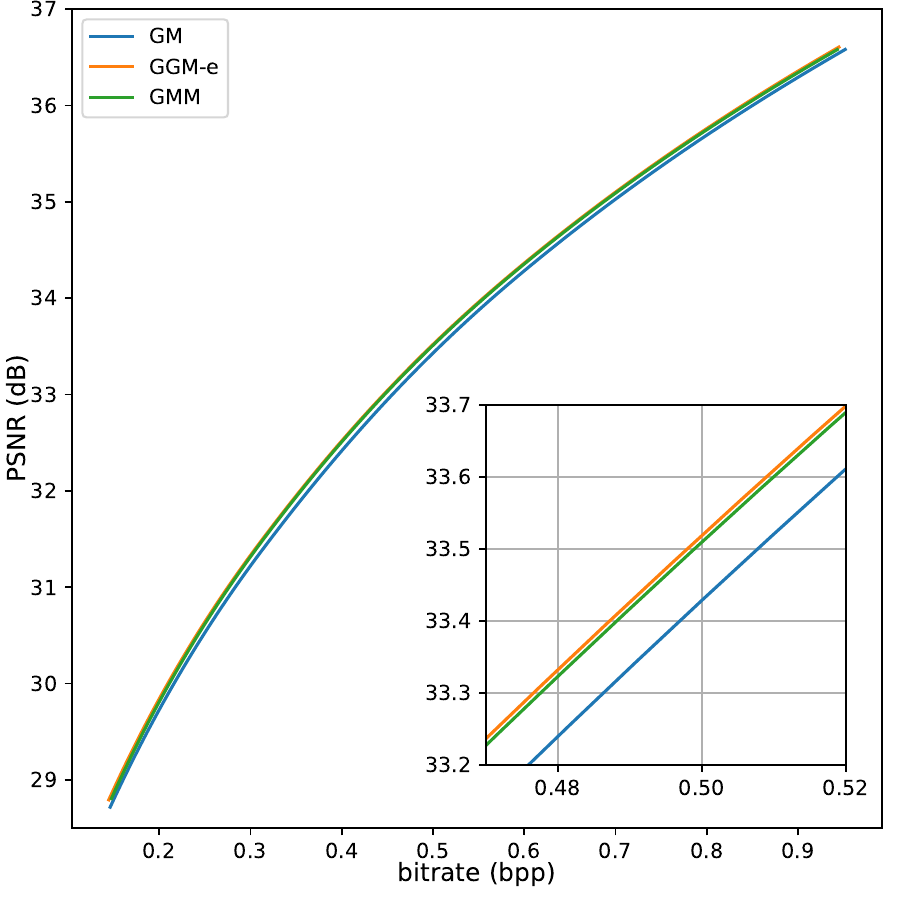}
        \label{fig:mse_single}
    }

    \subfloat[Charm\cite{minnen2020channel}]
    {  
        \includegraphics[width=0.24\linewidth]{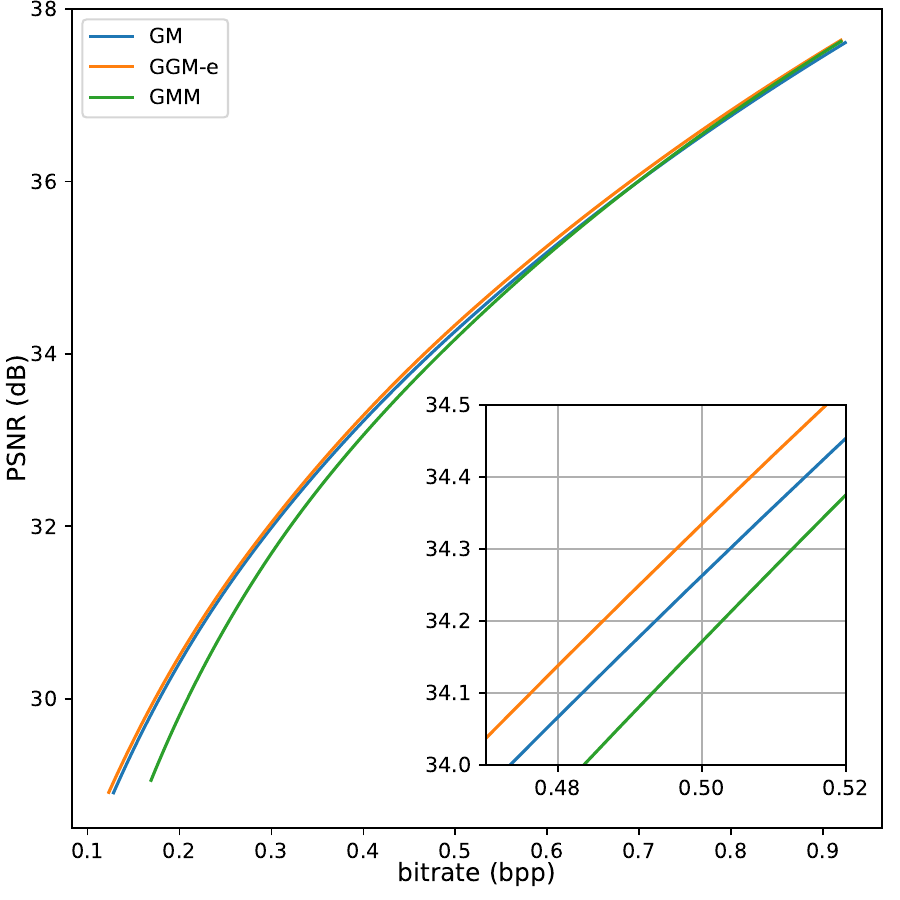}
        \label{fig:mse_single}
    } 
    \subfloat[ELIC\cite{he2022elic}]
    {  
        \includegraphics[width=0.24\linewidth]{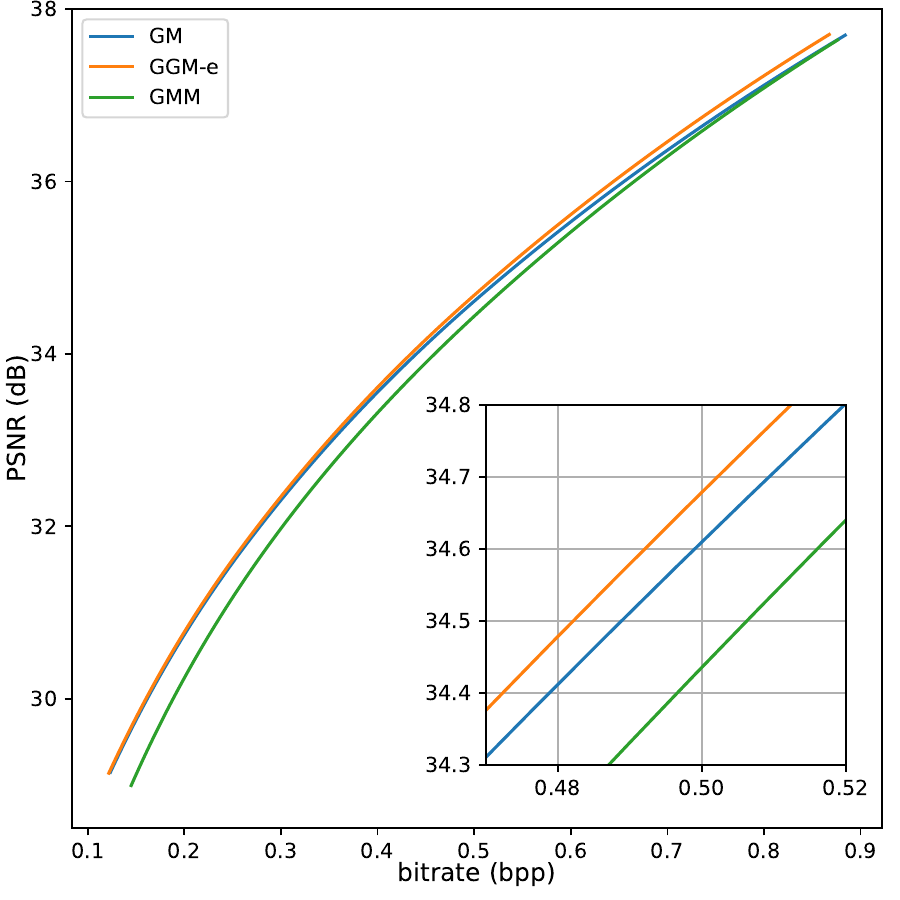}
        \label{fig:mse_single}
    } 
    \subfloat[TCM\cite{liu2023learned}]
    {  
        \includegraphics[width=0.24\linewidth]{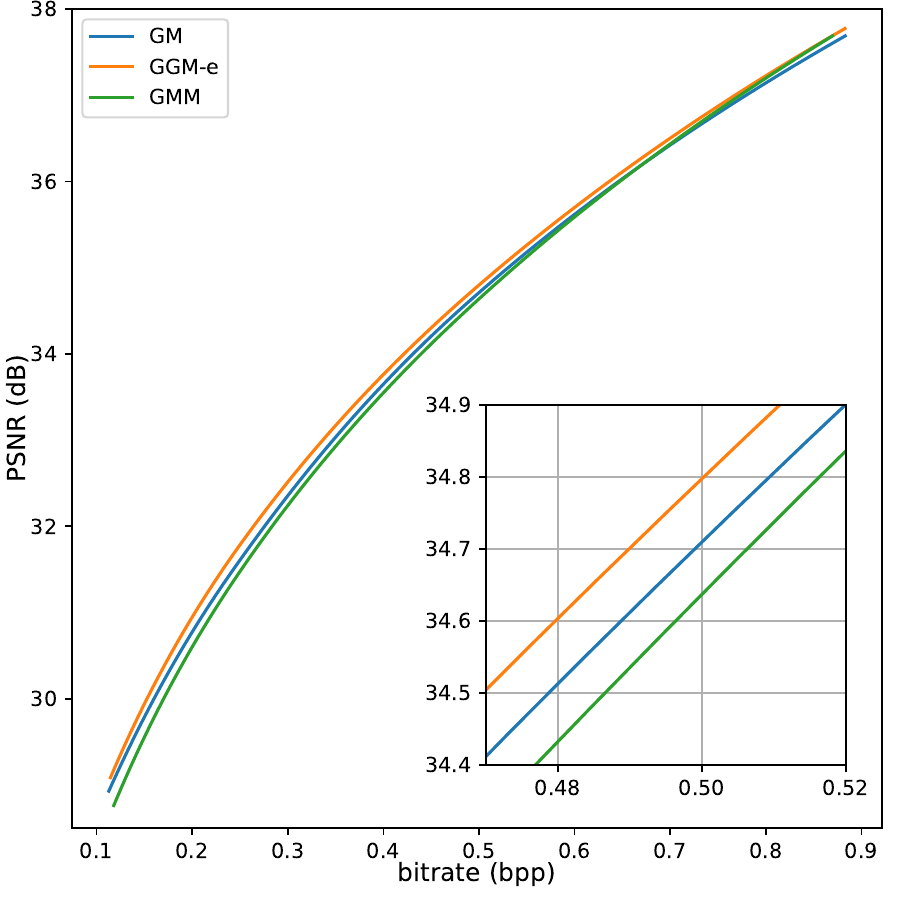}
        \label{fig:mse_single}
    } 
    \subfloat[FM-intra\cite{li2024fm}]
    {  
        \includegraphics[width=0.24\linewidth]{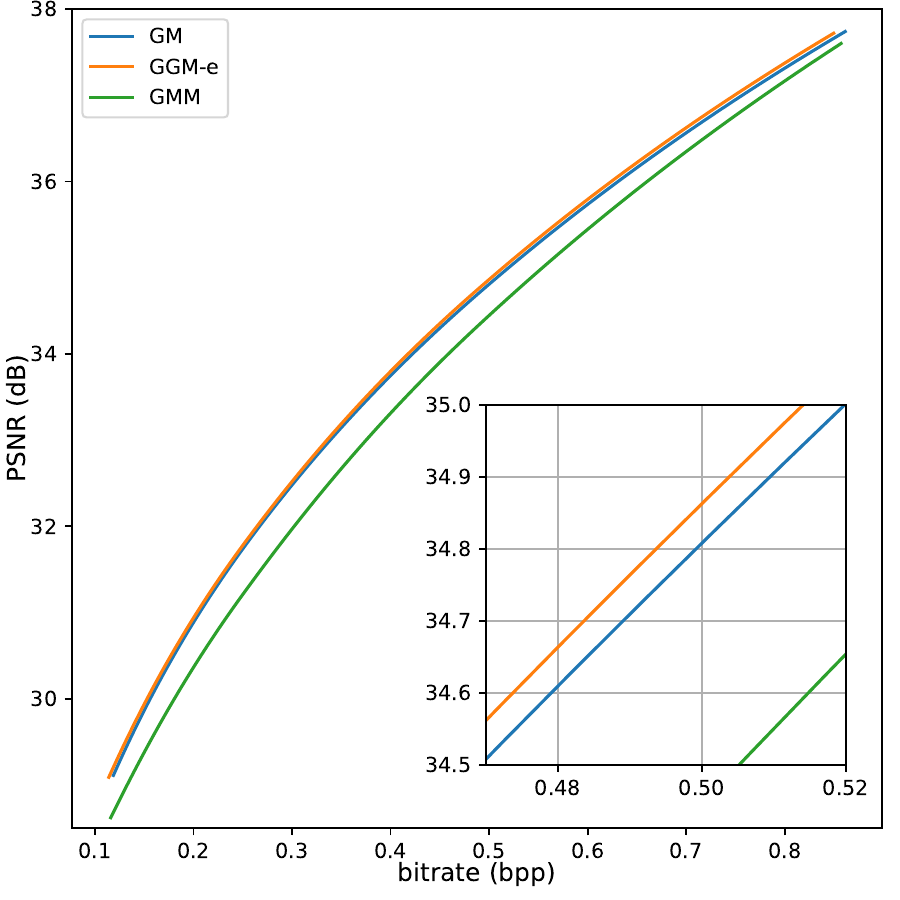}
        \label{fig:mse_single}
    } 
    
    \caption{
    Rate-distortion performance of utilizing GM, GMM, and our proposed GGM-e on various learned image compression methods. 
    }
    \label{fig:rd_curve}

\end{figure*}

\section{Experiments and Analyses}
\subsection{Experimental Setting}
\subsubsection{Learned image compression methods}
We verify the effectiveness of our methods on a variety of learned image compression methods:
\begin{itemize}
    \item \textbf{MS-hyper} (NeurIPS 2018) represents the mean-scale hyperprior model \cite{minnen2018joint}. The numbers of channels in latent space, $M$, and in transforms, $N$, are set as $M$=192 and $N$=128.
    \item \textbf{Charm} (ICIP 2020) denotes the channel-wise autoregressive model \cite{minnen2020channel}. We replace generalized division normalization layers with residual blocks as suggested in \cite{he2022elic} and change the kernel size in downsampling and upsampling layers from 5 to 4 as suggested in \cite{zhang2023practical}. The setting is $M$=320, $N$=192.
    \item \textbf{ELIC} (CVPR 2022) denotes the model proposed in \cite{he2022elic}. The kernel size in downsampling and upsampling layers is changed from 5 to 4. The setting is $M$=320, $N$=192. 
    \item  \textbf{Shallow-JPEG} and \textbf{Shallow-2layer} (ICCV 2023) denote models proposed in \cite{yang2023shallow}. We adopt the model with simpler analysis transform and non-residual connection synthesis transform for better performance as suggested in their open-source version\footnote{https://github.com/mandt-lab/shallow-ntc}.
    \item \textbf{TCM} (CVPR 2023) denote the model proposed in \cite{liu2023learned}. The setting is $N$=192.
    \item \textbf{FM-intra} (CVPR 2024) denotes the intra-coding model in the learned video compression model \cite{li2024fm}, which also achieves advanced performance. Note that the intra-coding model in learned video compression is a learned image compression model.
\end{itemize}
When implementing GGM-e and mixture models, we kept the same network structure in the entropy parameters module. The only difference is that the dimension of the output layer is higher for GGM-e and mixture models.

\begin{table*}[!t]
\setlength{\tabcolsep}{3.5pt}
\centering
\caption{Compression performance and complexity of different probabilistic models.}

\begin{threeparttable}
\begin{tabular}{ c|c|cc|ccc|ccc|ccc|ccc}
\hline
\multirow{3}{*}{Method}& \multirow{3}{*}{\Centerstack[c]{Prob.\\model}}  &  \multirow{3}{*}{\Centerstack[c]{Params\tnote{2}\\(M)}}  &  \multirow{3}{*}{\Centerstack[c]{KMACs\\per pixel}} & \multicolumn{3}{c|}{Kodak 512x768} & \multicolumn{3}{c|}{Tecnick 1200x1200} & \multicolumn{3}{c|}{CLIC 1200x1800} & \multicolumn{3}{c}{Avg.\tnote{3}}\\
&&&&\multirow{2}{*}{\Centerstack[c]{BD-\tnote{1}\\rate(\%)}}&\multirow{2}{*}{\Centerstack[c]{$t_e$\\(ms)}}&\multirow{2}{*}{\Centerstack[c]{$t_d$\\(ms)}}&\multirow{2}{*}{\Centerstack[c]{BD-\\rate(\%)}}&\multirow{2}{*}{\Centerstack[c]{$t_e$\\(ms)}}&\multirow{2}{*}{\Centerstack[c]{$t_d$\\(ms)}}&\multirow{2}{*}{\Centerstack[c]{BD-\\rate(\%)}}&\multirow{2}{*}{\Centerstack[c]{$t_e$\\(ms)}}&\multirow{2}{*}{\Centerstack[c]{$t_d$\\(ms)}}&\multirow{2}{*}{\Centerstack[c]{BD-\\rate(\%)}}&\multirow{2}{*}{\Centerstack[c]{$\Delta t_e$\\(\%)}}&\multirow{2}{*}{\Centerstack[c]{$\Delta t_d$\\(\%)}}\\
&&&&&&&&&&&&&&&\\
\hline
\multirow{5}{*}{\Centerstack[c]{Shallow-\\JPEG\cite{yang2023shallow}\\(ICCV 2023)}}&GM&10.5&19.1&0.00&42&41&0.00&144&138&0.00&232&208&0.00&0.00&0.00\\
&GMM&11.9&24.9&7.65&128&125&22.54&645&608&41.63&838&767&23.94&+271&+272\\
&GGM-m&10.5&19.1&-1.63&41&41&-1.16&144&139&-1.53&228&210&-1.44&-0.93&+0.55\\
&GGM-c&10.5&19.1&-1.87&44&42&-1.34&150&143&-1.52&235&222&\underline{-1.58}&+3.58&+4.51\\
&GGM-e&10.7&19.9&-2.91&44&43&-2.33&149&144&-3.15&240&219&\textbf{-2.80}&+4.27&+5.10\\
\hline
\multirow{5}{*}{\Centerstack[c]{Shallow-\\2layer\cite{yang2023shallow}\\(ICCV 2023)}}&GM&10.5&19.1&0.00&41&41&0.00&148&139&0.00&223&215&0.00&0.00&0.00\\
&GMM&11.9&24.9&3.45&138&125&9.77&644&608&10.04&834&769&7.75&+282&+267\\
&GGM-m&10.5&19.1&-1.02&40&40&-0.62&149&141&-0.70&225&213&-0.78&+0.05&-0.39\\
&GGM-c&10.5&19.1&-1.15&43&42&-1.08&151&148&-1.07&233&224&\underline{-1.10}&+3.76&+4.87\\
&GGM-e&10.7&19.9&-2.00&43&43&-1.67&151&148&-2.19&231&223&\textbf{-1.95}&+3.90&+5.35\\
\hline
\multirow{5}{*}{\Centerstack[c]{MS-\\hyper\cite{minnen2018joint}\\(NeurIPS 2018)}}&GM&4.0&7.2&0.00&26&28&0.00&128&94&0.00&218&144&0.00&0.00&0.00\\
&GMM&4.7&9.9&-1.98&75&69&-1.09&378&320&-1.20&578&465&\underline{-1.43}&+182&+201\\
&GGM-m&4.0&7.2&-0.63&26&29&-1.03&128&94&-0.81&219&143&-0.82&0.30&-0.18\\
&GGM-c&4.0&7.2&-1.64&28&30&-1.20&129&97&-0.92&223&150&-1.25&+3.09&+4.69\\
&GGM-e&4.3&8.2&-2.40&28&31&-2.07&130&98&-1.71&221&149&\textbf{-2.06}&+3.26&+4.90\\
\hline
\multirow{5}{*}{\Centerstack[c]{Charm\cite{minnen2020channel}\\(ICIP 2020)}}&GM&45.6&161.1&0.00&64&71&0.00&187&223&0.00&290&332&0.00&0.00&0.00\\
&GMM&54.7&187.8&8.64&149&143&10.14&815&856&10.01&952&1045&9.60&+233&+199\\
&GGM-m&45.6&161.1&-1.29&63&71&-0.97&189&223&-1.06&287&336&-1.11&-0.12&+0.27\\
&GGM-c&45.6&161.1&-1.42&69&75&-1.28&199&232&-1.43&301&346&\underline{-1.38}&+6.12&+4.18\\
&GGM-e&48.5&168.2&-1.79&68&76&-1.45&199&230&-1.61&302&343&\textbf{-1.62}&+6.16&+4.07\\
\hline
&GM&26.9&69.6&0.00&79&87&0.00&231&261&0.00&349&386&0.00&0.00&0.00\\
&GMM&27.8&73.5&7.63&179&174&10.31&911&928&10.36&1071&1145&9.43&+209&+184\\
&GGM-m&26.9&69.6&-0.52&78&87&-0.31&233&261&-0.21&347&389&-0.35&-0.10&+0.25\\
&GGM-c&26.9&69.6&-1.34&83&90&-0.86&243&269&-0.55&361&399&\underline{-0.92}&+5.01&+3.53\\
\multirow{-5}{*}{\Centerstack[c]{ELIC\cite{he2022elic}\\(CVPR 2022)}}&GGM-e&27.0&70.1&-1.72&83&91&-1.86&244&267&-2.15&362&396&\textbf{-1.91}&+5.04&+3.43\\
\hline
&GM&54.6&123.0&0.00&83&89&0.00&259&276&0.00&392&419&0.00&0.00&0.00\\
&GMM&56.9&137.8&3.51&144&134&5.18&645&616&5.05&917&874&4.58&+119&+94\\
&GGM-m&54.6&123.0&-1.19&87&93&-1.26&258&275&-1.36&392&419&-1.27&+1.43&+1.53\\
&GGM-c&54.6&123.0&-2.13&89&99&-1.92&261&283&-2.10&395&425&\underline{-2.05}&+2.91&+5.04\\
\multirow{-5}{*}{\Centerstack[c]{TCM\cite{liu2023learned}\\(CVPR 2023)}}&GGM-e&54.9&126.5&-2.76&87&97&-2.37&260&283&-2.57&396&426&\textbf{-2.57}&+2.33&+4.63\\
\hline
&GM&34.3&209.0&0.00&65&69&0.00&193&196&0.00&293&300&0.00&0.00&0.00\\
&GMM&36.2&224.0&11.63&161&152&13.13&783&784&13.39&982&946&12.72&+229&+212\\
&GGM-m&34.3&209.0&-0.89&65&68&-1.18&196&201&-1.04&294&303&-1.04&+0.95&+1.00\\
&GGM-c&34.3&209.0&-1.01&69&77&-1.24&199&208&-1.18&298&309&\underline{-1.15}&+3.97&+7.04\\
\multirow{-5}{*}{\Centerstack[c]{FM-\\intra\cite{li2024fm}\\(CVPR 2024)}}&GGM-e&34.5&212.2&-1.60&70&77&-1.63&199&208&-1.58&301&310&\textbf{-1.60}&+4.69&+7.26\\
\hline
\end{tabular}
\begin{tablenotes}
\footnotesize
\item[1] BD-rate$\downarrow$ is compared to the performance of GM. The average resolution of images in each dataset is presented. $t_e$ and $t_d$ represent the encoding and decoding time, respectively.
\item[3] Parameter count and KMACs/pixel only include the entropy model, since the analysis and synthesis transform have not been changed. We use the DeepSpeed library to evaluate these two metrics.
\item[4] In the `Avg.' column, the BD-rate is averaged across three datasets.
The best performance is marked in \textbf{bold}, and the second-best performance is \underline{underlined}. $\Delta t_e$ represents the increase in encoding time compared to GM. $\Delta t_d$ represents the increase in decoding time compared to GM. `+' indicates an increase in time, while `-' indicates a decrease in time. 
\end{tablenotes} 
\end{threeparttable}
\label{tab:main_performance}

\end{table*}

\subsubsection{Training} All models were optimized for Mean Squared Error (MSE). We trained multiple models with different values of $\lambda \in \{0.0018, 0.0054, 0.0162, 0.0483\}$. 
The training dataset is the Flicker2W \cite{liu2020unified} dataset, which contains approximately 20000 images. In each training iteration, images were randomly cropped into 256x256 patches. 
During training, we applied the Adam optimizer \cite{kingma2014adam} for 450 epochs, with a batch size of 8 and an initial learning rate of $10^{-4}$. After 400 epochs, the learning rate was reduced to $10^{-5}$, and after 30 epochs, it was further reduced to $10^{-6}$. For all of our experiments, we utilize the identical random seed. 

We use the mixed quantization surrogates for all models. If not explicitly stated, zero-center quantization ($\lfloor y-\mu \rceil+\mu$) is used for the Gaussian, Laplacian, Logistic, and generalized Gaussian model, where the mean parameter is clearly identifiable, as suggested in \cite{minnen2020channel,zhang2023uniform}. Following \cite{cheng2020learned,fu2023learned,mentzer2023m2t}, we use nonzero-center quantization ($\lfloor y \rceil$) for these mixture models, such as GMM and GLLMM, since there is not a clearly identifiable mean parameter. The studies before \cite{zhang2023uniform} did not give much consideration to the effect of lower bound for scale parameters. Zhang \textit{et al.} \cite{zhang2023uniform} explored the appropriate lower bounds for the scale parameters of Gaussian and Laplacian.
In this paper, we empirically find proper bounds for scale parameters of these probabilistic models we compared. For GM, the bound is set as 0.11. For the Laplacian model (LaM), the bound is set as 0.06. For the Logistic model (LoM), the bound is set as 0.04. Each component in GMM and GLLMM uses its corresponding optimal bound. The experimental results for determining the proper bound are included in Appendix \ref{supp:7}.
During the training of GGM, we restrict the range of $\beta$ in $[0.5,4]$ to avoid the numerical computation error.

\subsubsection{Entropy coding}
For GM and GGM, we follow the previous studies \cite{balle2019integer,sun2021learned, begaint2020compressai} to adopt the look-up tables-based (LUTs-based) method for entropy coding, as introduced in Sec. \ref{sec:lut}. In contrast, for GMM, which has 9 parameters, using the LUTs-based method would incur excessive storage costs. For example, sampling 20 values for each of the 9 parameters in GMM would require storing $20^9$ CDF tables, resulting in a storage requirement of at least $2 \times 10^5$ GB. Therefore, following previous studies \cite{cheng2020learned, fu2023learned}, we calculate the CDF tables dynamically during encoding and decoding for GMM. Details about entropy coding are included in Appendix \ref{supp:5} and \ref{supp:6}.

\subsubsection{Evaluation}
We evaluate various methods on three commonly used test datasets. The Kodak dataset \footnote{Available online at \url{http://r0k.us/graphics/kodak/}.} contains 24 images with either 512×768 or 768×512 pixels. The Tecnick dataset \cite{asuni2014testimages} contains 100 images with 1200×1200 pixels. The CLIC 2020 professional validation dataset\footnote{http://compression.cc}, which we refer to as CLIC in this paper, includes 41 high-quality images. 

We use bits-per-pixel (bpp) and Peak Signal-to-Noise Ratio (PSNR) to quantify rate and distortion.
The PSNR is measured in the RGB space. 
We also use the BD-Rate metric \cite{bjontegaard2001calculation} to calculate the rate saving. To obtain the overall BD-Rate for an entire dataset, we first calculate the BD-Rate of each image and then average them over all images. 
The algorithm's runtime was tested on a single-core Intel(R) Xeon(R) Gold 6248R CPU @ 3.00GHz and one NVIDIA GeForce RTX 3090 GPU.
We test the performance of BPG-0.9.8\footnote{https://bellard.org/bpg/} and VTM-22.0\footnote{https://vcgit.hhi.fraunhofer.de/jvet/VVCSoftware\_VTM} with input format as YUV444.

\begin{figure}[!t]
    \centering
  \subfloat[GGM-m]
    {  
        \includegraphics[width=0.95\linewidth]{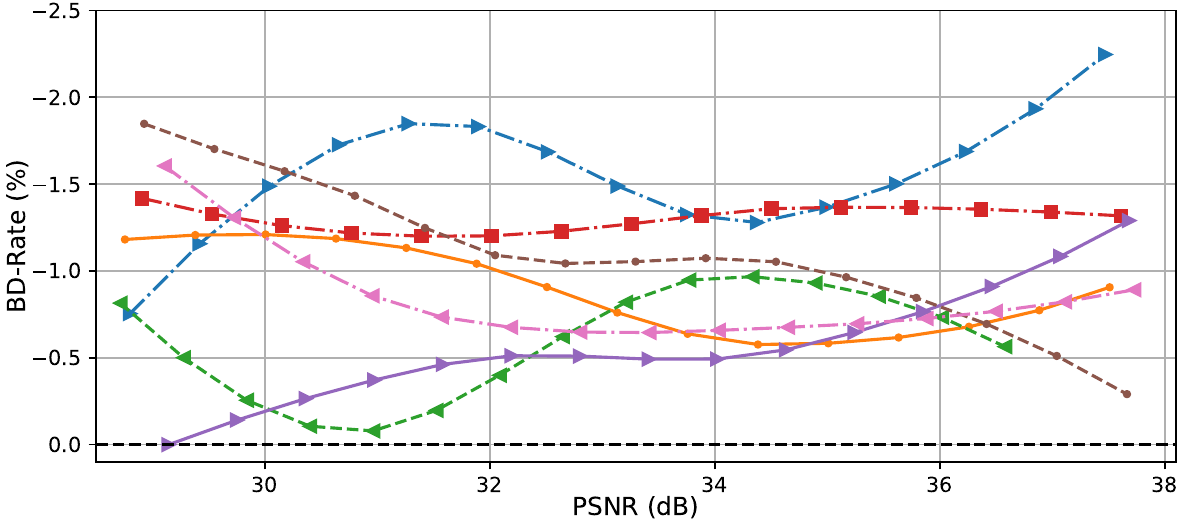}
        \label{fig:mse_single}
    }

    \subfloat[GGM-c]
    {  
        \includegraphics[width=0.95\linewidth]{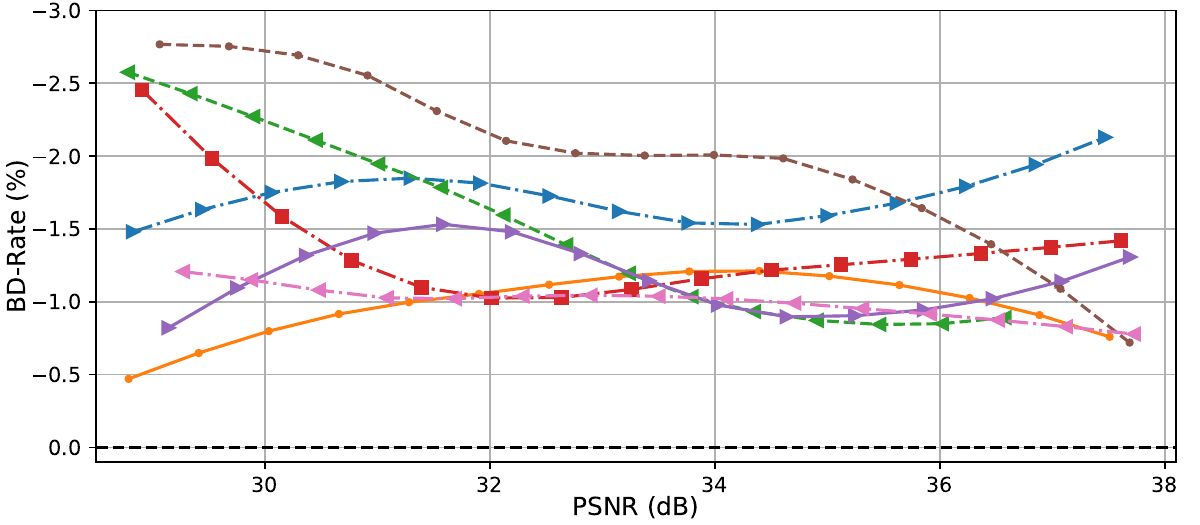}
        \label{fig:mse_single}
    }

    \subfloat[GGM-e]
    {  
        \includegraphics[width=0.95\linewidth]{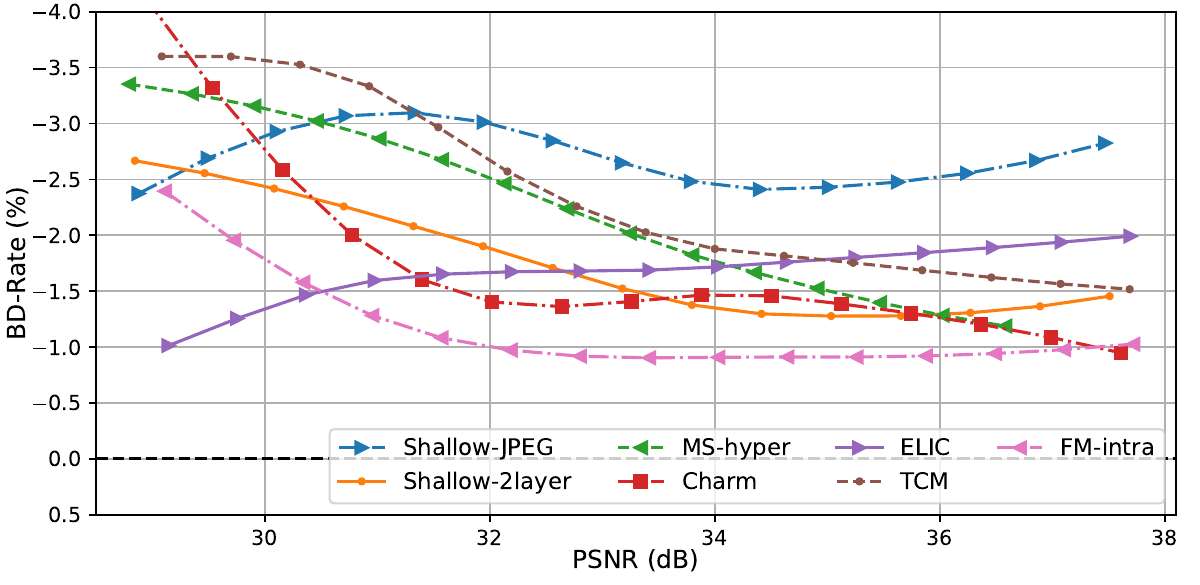}
        \label{fig:mse_single}
    } 
    
    \caption{
    BD-Rate$\downarrow$ (\%) of GGM compared to GM on the Kodak dataset at different bitrate ranges.
    }
    \label{fig:vary_psnr}

\end{figure}

\subsection{Performance}
\subsubsection{Compression performance}
Table \ref{tab:main_performance} reports the performance of GM, GMM, and our GGM methods on various learned image compression methods. Figure \ref{fig:rd_curve} shows the rate-distortion curves with different probabilistic models. Figure \ref{fig:vary_psnr} further shows the performance improvement at different bitrate ranges.

Table \ref{tab:main_performance} shows that our GGM-m, GGM-c, and GGM-e outperform GM on a variety of learned image compression models. From GGM-m to GGM-c to GGM-e, as the diversity of $\beta$ increases, the distribution modeling capacity improves, resulting in better compression performance.
Our GGM-e also performs better than GMM on all image compression models we evaluated.
The GMM performs poorly on some image compression models due to the severe influence of train-test mismatch caused by nonzero-center quantization \cite{zhang2023uniform}, especially at a lower bitrate. This phenomenon will be discussed in Sec. \ref{sec:others}.
Moreover, as shown in Fig. \ref{fig:vary_psnr}, the performance improvement brought by GGM differs across the bitrate ranges. The reason will be discussed in Sec. \ref{sec:distribution}.

On the recent advanced compression methods, our GGM also achieves notable performance improvement. For ELIC, GGM-m, GGM-c, and GGM-e achieve 0.35\%, 0.92\%, and 1.91\% rate savings, respectively, compared to GM. For TCM, GGM-m, GGM-c, and GGM-e achieve 1.27\%, 2.05\%, and 2.57\% rate savings, respectively. For FM-intra, GGM-m, GGM-c, and GGM-e achieve 1.04\%, 1.15\%, and 1.60\% rate savings, respectively.

\subsubsection{Complexity}
The network complexity and actual coding time are summarized in Table \ref{tab:main_performance}. Since the GGM-m model has only one $\beta$, similar to GM, its network complexity and coding time are comparable to those of GM. For GGM-c, the network complexity is also similar to GM, 
while its coding time is longer due to the increased number of CDF tables required for entropy coding. The increased number of CDF tables leads to a longer time for generating indexes of CDF tables, as well as longer memory access time. For GGM-e, the network complexity and coding time are both higher than GM, primarily due to the larger output dimension of entropy parameters in the entropy model and the additional CDF tables for entropy coding. The increases in coding time for GGM-c and GGM-e are similar, remaining under 8\% compared to GM.

For GMM, the CDF tables need to be dynamically generated for each latent element. Moreover, since the number of CDF tables for GMM equals the number of latent variables, which is significantly larger than that in the LUTs-based approach for GM and GGM, the time required for memory access becomes a notable factor. For instance, in the ELIC model, with an image resolution of $512 \times 768$, approximately 0.47 million CDF tables need to be calculated, resulting in a memory cost of 480MB when each CDF table has a length of 256. In contrast, GGM-c and GGM-e require only 1.56MB of memory. Thus, the memory access time for GMM is considerably higher. Due to the increased network complexity and the additional time costs associated with calculating CDF tables and memory access, the coding time for GMM is significantly longer than that for GM.

\subsection{Analyses on the Distribution of Shape Parameters}
\label{sec:distribution}
In this section, we analyze the distribution of learned shape parameters to better understand the effectiveness of GGM.

\begin{table}[!t]
\centering
\caption{Learned model-wise shape parameter $\beta$ in different models.}

\begin{threeparttable}
\begin{tabular}{ cccccc} 
\hline
\Centerstack[c]{Method} & \Centerstack[c]{$\lambda=$\\$0.0018$} & \Centerstack[c]{$\lambda=$\\$0.0054$} &\Centerstack[c]{$\lambda=$\\$0.0162$} & \Centerstack[c]{$\lambda=$\\$0.0483$}&Avg.\\
\hline
Shallow-JPEG\cite{yang2023shallow}&1.40 &1.30 &1.27 &1.29&1.32 \\
Shallow-2layer\cite{yang2023shallow}&1.47 &1.33 &1.34 &1.42&1.39 \\
MS-hyper\cite{minnen2018joint}&1.36 &1.38 &1.41 &1.45&1.40 \\
Charm\cite{minnen2020channel}&1.46 &1.45 &1.37 &1.39&1.42 \\
ELIC\cite{he2022elic}&1.82 &1.66 &1.59 &1.50&1.64 \\
TCM\cite{liu2023learned}&1.68 &1.76 &1.70 &1.84&1.74 \\
FM-intra\cite{li2024fm}&1.65 &1.63 &1.55 &1.58&1.60 \\
\hline
\end{tabular}
\begin{tablenotes}
\footnotesize
\item[1] Larger $\lambda$ represents higher bitrate. `Avg.' represents the average value across four models targeting different bitrates.
\end{tablenotes} 
\end{threeparttable}

\label{tab:detail_ggm_m_beta}
\end{table}

\begin{figure}[!t]
\centering
\subfloat[MS-hyper\cite{minnen2018joint} $\lambda=0.0018$]
{  
    \includegraphics[width=0.46\linewidth]{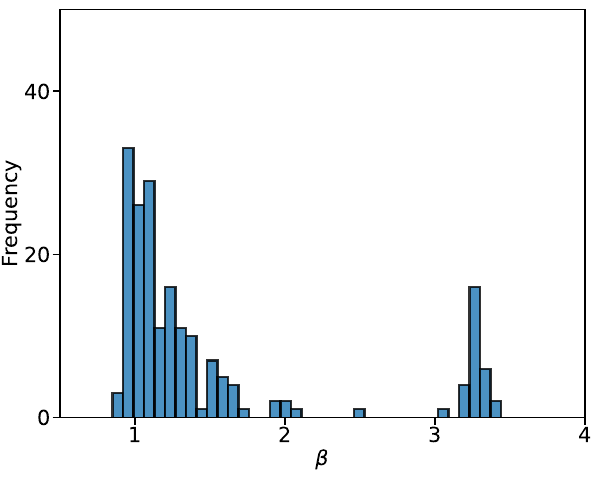}
}
\subfloat[MS-hyper\cite{minnen2018joint} $\lambda=0.0483$]
{  
    \includegraphics[width=0.46\linewidth]{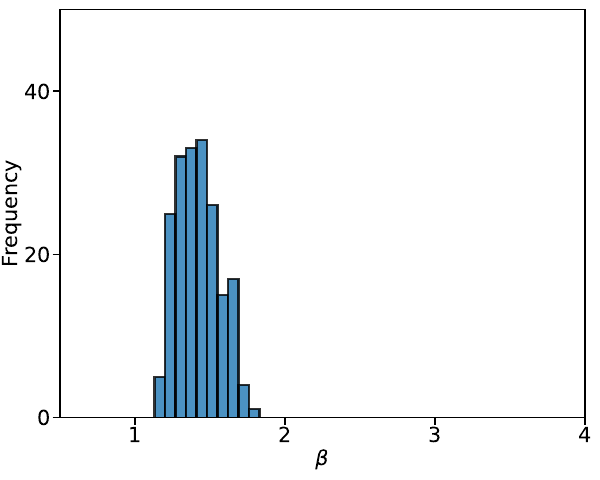}
}

\subfloat[ELIC\cite{he2022elic} $\lambda=0.0018$]
{  
    \includegraphics[width=0.46\linewidth]{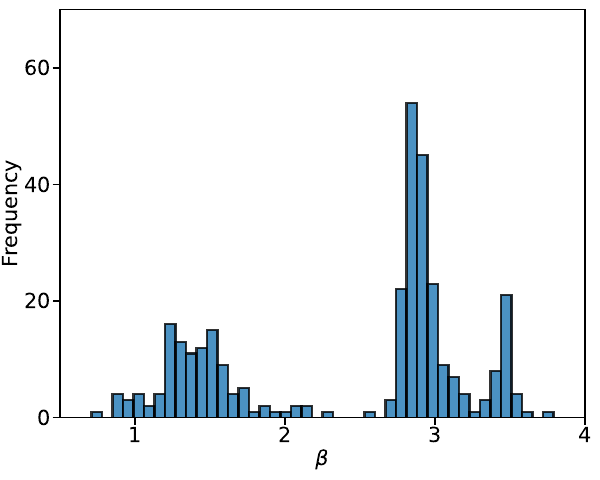}
}
\subfloat[ELIC\cite{he2022elic} $\lambda=0.0483$]
{  
    \includegraphics[width=0.46\linewidth]{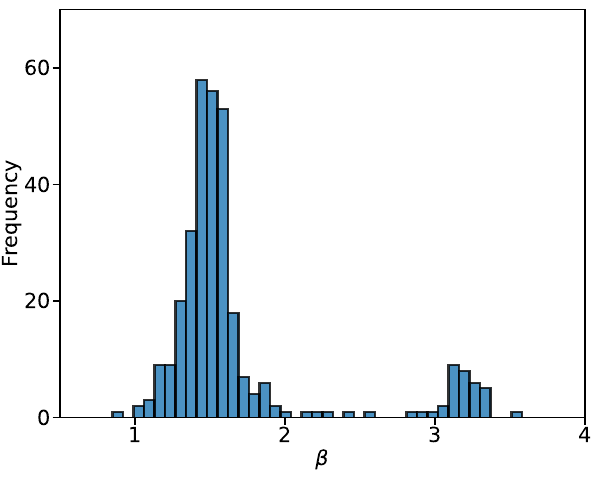}
} 
\caption{
Distribution of channel-wise shape parameters ($\beta$) in different learned image compression models. For MS-hyper, there are 192 channels, and for ELIC, there are 320 channels.
}
\label{fig:ggm_c}

\end{figure}

For GGM-m, detailed values of learned model-wise shape parameters are presented in Table \ref{tab:detail_ggm_m_beta}. Since the characteristics of latent variables differ across learned image compression models and bitrates, the learned model-wise shape parameters vary across models and bitrate ranges. However, all values are distinct from 2. This suggests that the Gaussian model ($\beta=2$) is insufficient to accurately fit the distribution, while a generalized Gaussian model with a different $\beta$ provides a better result. For GGM-c, the distribution of channel-wise shape parameters is shown in Fig. \ref{fig:ggm_c}. The channel-wise shape parameters are not the same as the model-wise shape parameters, showing the effectiveness of channel-wise shape parameters.

\begin{figure}[!t]
\centering
\subfloat[MS-hyper\cite{minnen2018joint} $\lambda=0.0018$]
{  
    \includegraphics[width=0.46\linewidth]{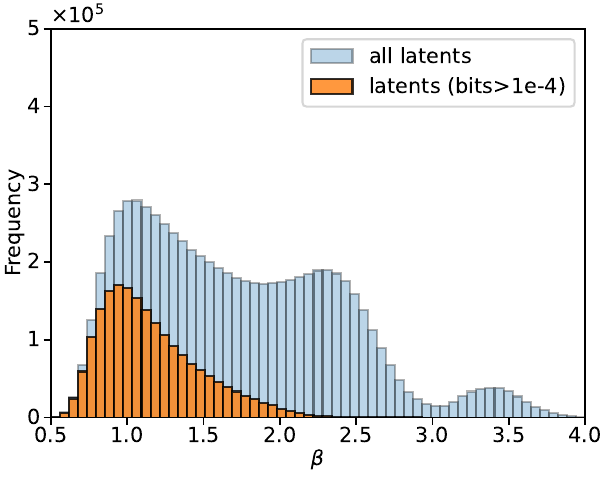}
}
\subfloat[MS-hyper\cite{minnen2018joint} $\lambda=0.0483$]
{  
    \includegraphics[width=0.46\linewidth]{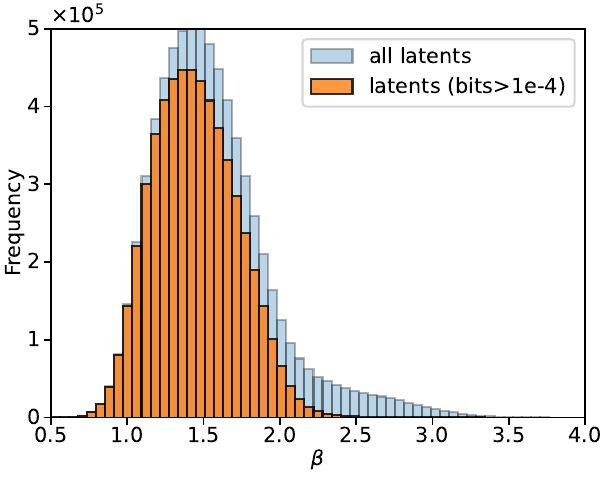}
}

\subfloat[ELIC\cite{he2022elic} $\lambda=0.0018$]
{  
    \includegraphics[width=0.46\linewidth]{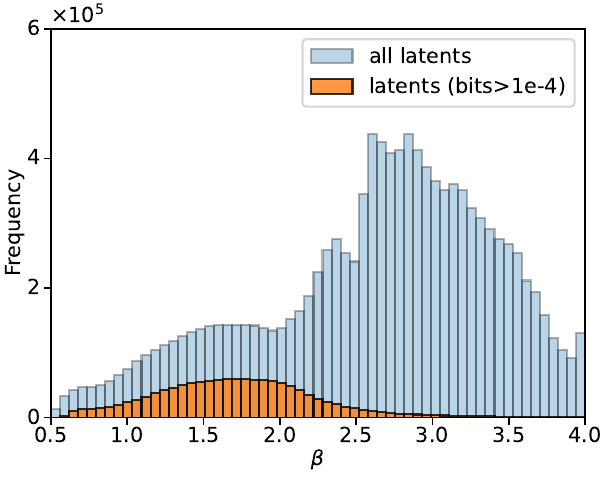}
}
\subfloat[ELIC\cite{he2022elic} $\lambda=0.0483$]
{  
    \includegraphics[width=0.46\linewidth]{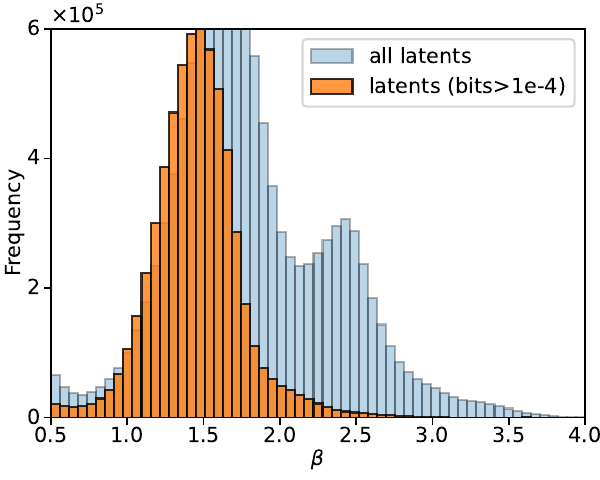}
} 
\caption{
Distribution of element-wise shape parameters ($\beta$) in different learned image compression models. The blue bars represent the distribution of $\beta$ for all latent variables, while the orange bars represent the distribution for latent variables with bits larger than $1 \times 10^{-4}$. The samples are collected from the Kodak dataset. For MS-hyper, there are 7,077,888 samples, and for ELIC, there are 11,796,480 samples.
}
\label{fig:ggm_e}

\end{figure}

For GGM-e, we have provided the distribution of the shape parameters, as shown in Fig. \ref{fig:ggm_e}. Since the distribution characteristics of latent variables differ across learned image compression models and bitrate ranges, the distribution of element-wise shape parameters also varies accordingly. In learned image compression models, many latent variables with extremely low bitrates contribute little to the overall rate. The accuracy of distribution modeling for these latent variables is less crucial. Consequently, to better understand the distribution of element-wise shape parameters, we present the distribution of shape parameters for latent variables with bits larger than $1\times 10^{-4}$. As shown in Fig. \ref{fig:ggm_e}, most of the shape parameters for these latent variables are concentrated between 0.5 and 2.5 and are distinct from the model-wise shape parameters. This indicates that different latent variables correspond to different shape parameters, demonstrating the effectiveness of the element-wise shape parameters.

\begin{table}[!t]
\centering
\caption{Average of element-wise shape parameters in different models.}
\begin{threeparttable}
\begin{tabular}{ ccccc} 
\hline
\Centerstack[c]{Method} & \Centerstack[c]{$\lambda=0.0018$} & \Centerstack[c]{$\lambda=0.0054$} &\Centerstack[c]{$\lambda=0.0162$} & \Centerstack[c]{$\lambda=0.0483$}\\
\hline
MS-hyper&1.16 &1.24 &1.37 &1.45 \\
ELIC&1.73 &1.62 &1.59 &1.45 \\
\hline
\end{tabular}
\begin{tablenotes}
\footnotesize
\item[1] Larger $\lambda$ represents higher bitrate. We only average the shape parameters of these latent variables with bits larger than 1e-4.
\end{tablenotes} 
\end{threeparttable}
\label{tab:detail_ggm_e_beta}

\end{table}

Moreover, we analyze the phenomenon that the performance improvement brought by GGM differs across bitrate ranges by analyzing the distribution of $\beta$ parameters. We use MS-hyper \cite{minnen2018joint} and ELIC \cite{he2022elic} as examples to analyze the performance gain of applying GGM-e across different bitrate ranges. As shown in Fig. \ref{fig:vary_psnr}c, on average, applying GGM-e performs better at relatively lower bitrates for MS-hyper, while it yields better performance at higher bitrates for ELIC.
The performance gain brought by GGM comes from the flexibility introduced by the additional shape parameter. If the distribution of latent variables in the learned image compression model is already well-fitted by the Gaussian model ($\beta = 2$), the performance improvement of GGM will be limited. Since latent variables with extremely low bitrates contribute little to the overall rate, the accuracy of distribution modeling for these variables is not crucial. Therefore, we focus on the distribution of element-wise shape parameters for latent variables with bits greater than $1\times 10^{-4}$, which is shown in Fig. \ref{fig:ggm_e}.

\begin{table}[!t]
    \centering
    \caption{BD-Rate$\downarrow$ (\%) of various probabilistic models\tnote{1}.}
    
    \begin{threeparttable}
    \begin{tabular}{ c|cc|cc } 
    \hline
    \Centerstack[c]{quantization\\method}  & \Centerstack[c]{probabilistic\\model} & n\tnote{2}&\Centerstack[c]{proper\tnote{3}\\scale bound} & \Centerstack[c]{tiny\tnote{4}\\scale bound} \\
    \hline
\multirow{4}{*}{\Centerstack[c]{zero-center\\$\lfloor y-\mu\rceil+\mu$}}&GM&2&0.00\tnote{5}&4.17\\
&LaM&2&0.25&4.73\\
&LoM&2&-0.35&\underline{4.10}\\
&GGM-e&3&\textbf{-2.40}&\textbf{2.80}\\
\hline
\multirow{6}{*}{\Centerstack[c]{nonzero-center\\$\lfloor y\rceil$}}&GM&2&0.67&6.98\\
&LaM&2&0.51&8.68\\
&LoM&2&0.56&8.12\\
&GGM-e&3&-1.47&5.82\\
&GMM&9&\underline{-1.98}&6.05\\
&GLLMM&30&-0.93&5.92\\
    \hline
    \end{tabular}
    \begin{tablenotes}
    \footnotesize
    \item[1] The results are tested with MS-hyper \cite{minnen2018joint} model on Kodak dataset. The best performances in each column are marked in \textbf{bold}, while the second best are \underline{underlined}.
    \item[2] $n$ represents the number of parameters that need to be estimated.
    \item[3] For GGM, we use our proposed $\beta$-dependent lower bound for scale parameters. For others, we empirically find the optimal lower bound.
    \item[4] The value of the tiny lower bound is $10^{-6}$, which is set to prevent the scale parameter from approaching zero.
    \item[5] The performance of GM with zero-center quantization and proper scale bound is set as the anchor for calculating BD-Rate.
\end{tablenotes} 
    \end{threeparttable}
    \label{tab:probabilistic_other}
    
\end{table}

\begin{table}[!t]
\centering
\caption{Performance of different quantization methods on various models.}

\begin{threeparttable}
\begin{tabular}{cccc}
\hline
\multirow{2}{*}{Method} & \multirow{2}{*}{\Centerstack[c]{Quantization\\method}} & \multirow{2}{*}{\Centerstack[c]{Probabilistic\\model}}  & \multirow{2}{*}{\Centerstack[c]{BD-rate (\%)}} \\
&&&\\
\hline
\multirow{3}{*}{\Centerstack[c]{MS-hyper}}&$\lfloor y-\mu \rceil+\mu$&GM&0.00\\
\cline{2-4}
&\multirow{2}{*}{\Centerstack[c]{$\lfloor y \rceil$}}&GM&0.67\\
&&GMM&-1.98\\
\hline
\multirow{3}{*}{\Centerstack[c]{Shallow-\\JPEG}}&$\lfloor y-\mu \rceil+\mu$&GM&0.00\\
\cline{2-4}
&\multirow{2}{*}{\Centerstack[c]{$\lfloor y \rceil$}}&GM&11.22\\
&&GMM&7.65\\
\hline
\end{tabular}
\begin{tablenotes}
\footnotesize
\item[1] BD-rate$\downarrow$ is compared to the performance of GM with zero-center quantization $\lfloor y-\mu \rceil+\mu$ on the Kodak dataset.
\end{tablenotes} 
\end{threeparttable}
\label{tab:quantization_offset}

\end{table}

As shown in Fig. \ref{fig:ggm_e}, in MS-hyper, the shape parameters are concentrated around 1.0 at lower bitrates and around 1.5 at higher bitrates. The average values of these shape parameters are presented in Table \ref{tab:detail_ggm_e_beta}. As shown in Table \ref{tab:detail_ggm_e_beta}, the average shape parameter is farther from 2 at lower bitrates compared to higher bitrates. This suggests that the distribution of latent variables in MS-hyper deviates more from a Gaussian distribution at lower bitrates, resulting in a higher performance gain when applying GGM-e at lower bitrates. In contrast, for the ELIC model, the shape parameters are concentrated around 1.75 at lower bitrates and around 1.5 at higher bitrates. As shown in Table \ref{tab:detail_ggm_e_beta}, the average shape parameter is farther from 2 at higher bitrates than at lower bitrates. This indicates that the latent variable distributions in ELIC deviate more from a Gaussian distribution at higher bitrates, leading to a more significant performance gain when applying GGM-e at higher bitrates.

\subsection{Comparison with Other Probabilistic Models}
\label{sec:others}
We compare the performance of our GGM-e method with other commonly used probabilistic models. Table \ref{tab:probabilistic_other} reports the results of various probabilistic models with different quantization methods and different settings of lower bound for the scale parameter. As shown in Table \ref{tab:probabilistic_other}, a proper lower bound for scale parameters can enhance the performance of all these probabilistic models. 
Using zero-center quantization further improves the performance of these distributions with a clearly identifiable mean parameter, such as GM, LaM, LoM, and GGM. With the help of zero-center quantization, GGM-e outperforms GMM.

As shown in Table \ref{tab:main_performance}, among the learned image compression models we tested, applying GMM only on MS-hyper outperforms GM. For all other models, GMM performs worse than GM. We further evaluated the influence of different quantization methods on other learned image compression models. The experimental results are shown in Table \ref{tab:quantization_offset}. For MS-hyper, applying $\lfloor y \rceil$ with GM results in slightly worse performance compared to $\lfloor y-\mu \rceil + \mu$. For Shallow-JPEG, however, applying $\lfloor y \rceil$ with GM performs significantly worse than $\lfloor y-\mu \rceil + \mu$. This performance gap arises from the train-test mismatch introduced by quantization approximation during training, and the effect of this mismatch varies across different learned image compression models \cite{zhang2023uniform}. When comparing GM and GMM using the same quantization method $\lfloor y \rceil$, GMM outperforms GM.

\begin{figure*}[!t]
    \centering
  \subfloat[MS-hyper GGM-m]
    {  
        \includegraphics[width=0.48\linewidth]{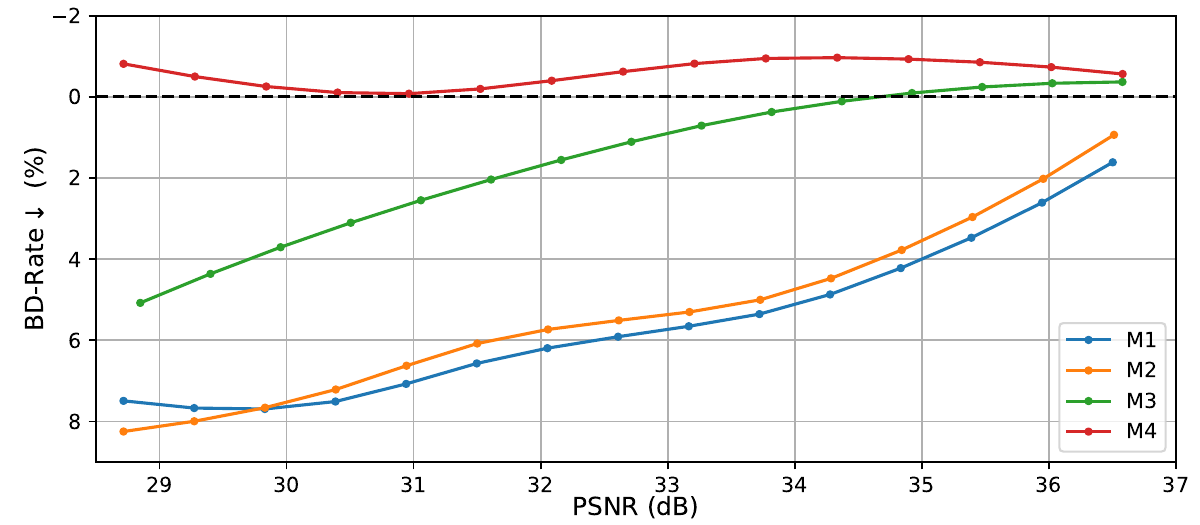}
        \label{fig:mse_single}
    } 
    \subfloat[MS-hyper GGM-c]
    {  
        \includegraphics[width=0.48\linewidth]{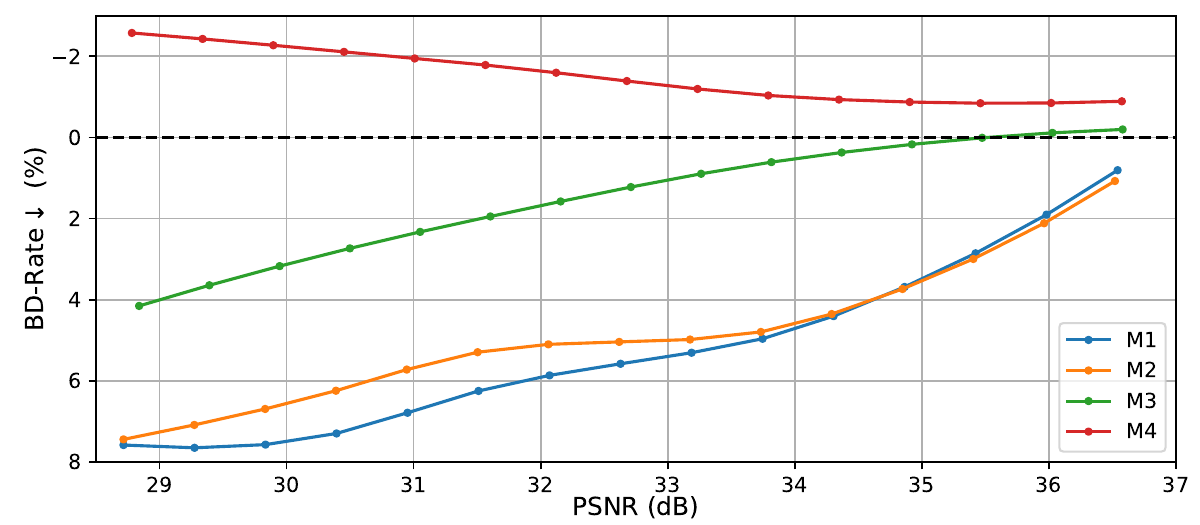}
        \label{fig:mse_single}
    }

    \subfloat[MS-hyper GGM-e]
    {  
        \includegraphics[width=0.48\linewidth]{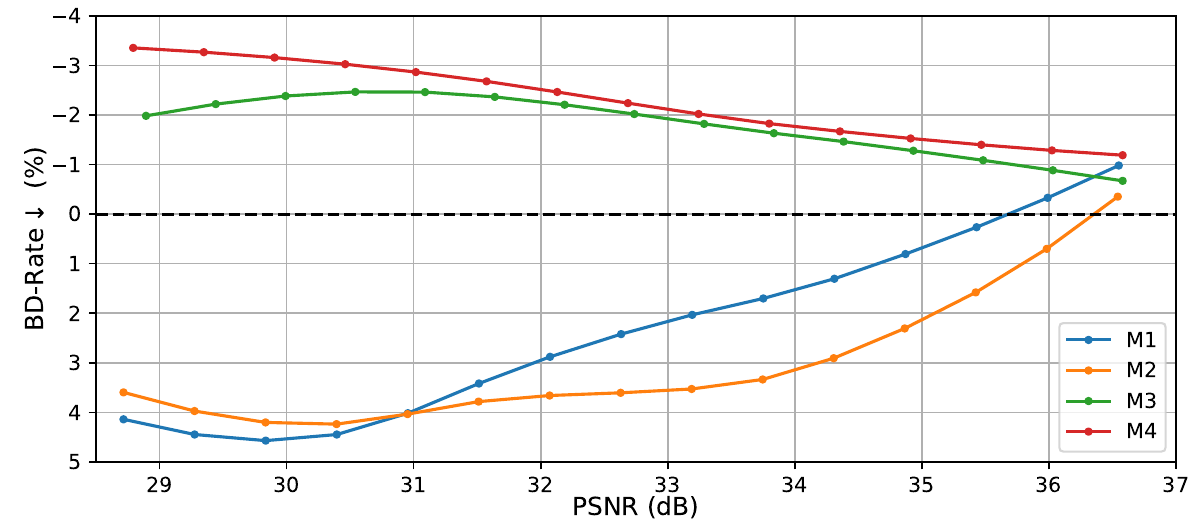}
        \label{fig:mse_single}
    } 
    \subfloat[ELIC GGM-e]
    {  
        \includegraphics[width=0.48\linewidth]{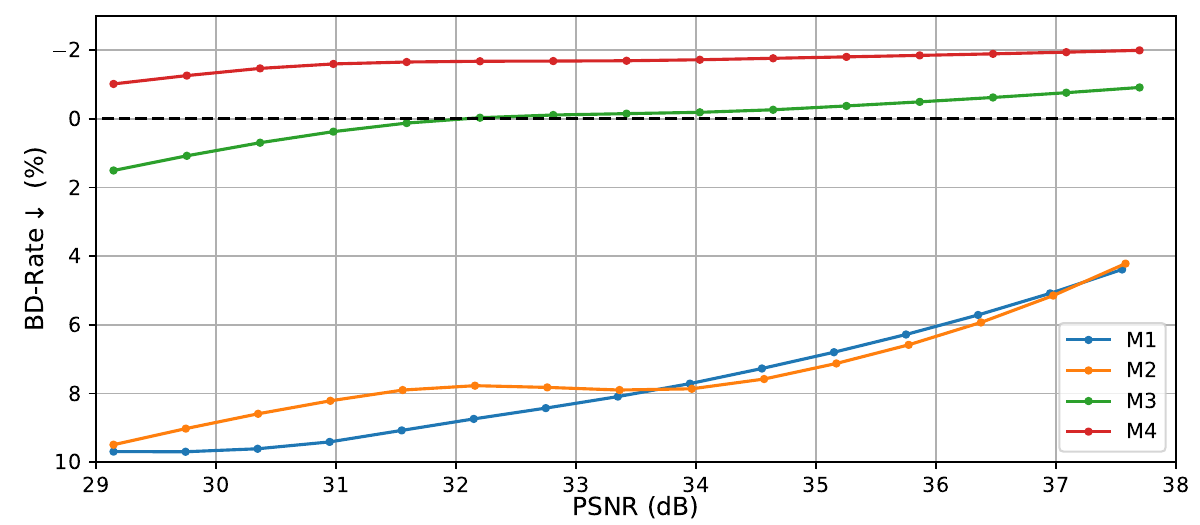}
        \label{fig:mse_single}
    } 
    
    \caption{
    BD-Rate$\downarrow$ (\%) at different bitrate ranges of ablation studies for improved training methods. The notations (M1, M2, M3, M4) refer to Table \ref{tab:ablation}. The BD-Rate is averaged on the Kodak dataset.
    }
    \label{fig:ablation}

\end{figure*}
\subsection{Ablation Studies}
We conduct a series of ablation studies to verify the contribution of each component in our proposed improved training methods. Table \ref{tab:ablation} shows that the $\beta$-dependent lower bound method significantly improves the performance (M1$\rightarrow$M3). The gradient rectification method could further improve performance based on $\beta$-dependent lower bound (M3$\rightarrow$M4). Only using the gradient rectification has little influence on the performance (M1$\rightarrow$M2). Figure \ref{fig:ablation} further shows the rate saving at different bitrate ranges.
\begin{table}[!t]
    \centering
    \caption{BD-Rate$\downarrow$ (\%) of ablation studies on $\beta$-dependent lower bound and gradient rectification.}
    \tabcolsep=4pt
    
    \begin{threeparttable}
    \begin{tabular}{ccc|ccc|c} 
    \hline
    &\multirow{2}{*}{\Centerstack[c]{$\beta$-dependent\\lower bound}} & \multirow{2}{*}{\Centerstack[c]{gradient\\rectification}}  &\multicolumn{3}{c|}{MS-hyper\tnote{1}}&ELIC\tnote{2} \\
      && & \Centerstack[c]{GGM-m} & \Centerstack[c]{GGM-c} & GGM-e&GGM-e \\
    \hline
M1&\usym{2717}&\usym{2717}&6.77&6.40&2.87&7.44\\
M2&\usym{2717}&\usym{2713}&6.50&5.92&3.78&7.52\\
M3&\usym{2713}&\usym{2717}&1.48&1.38&-1.93&-0.69\\
M4&\usym{2713}&\usym{2713}&\textbf{-0.63}&\textbf{-1.64}&\textbf{-2.40}&\textbf{-1.72}\\
    \hline
    \end{tabular}
    \begin{tablenotes}
    \footnotesize
    \item[1,2] The anchor of BD-Rate in this table is the performance of GM with a proper scale lower bound of the corresponding image compression method. The BD-Rate is averaged on the Kodak dataset.
\end{tablenotes} 
    \end{threeparttable}
    \label{tab:ablation}
    
\end{table}

\begin{figure*}[!t]
    \centering
  \subfloat[MS-hyper GGM-c $\lambda$=0.0018]
    {  
    \includegraphics[width=0.49\linewidth]{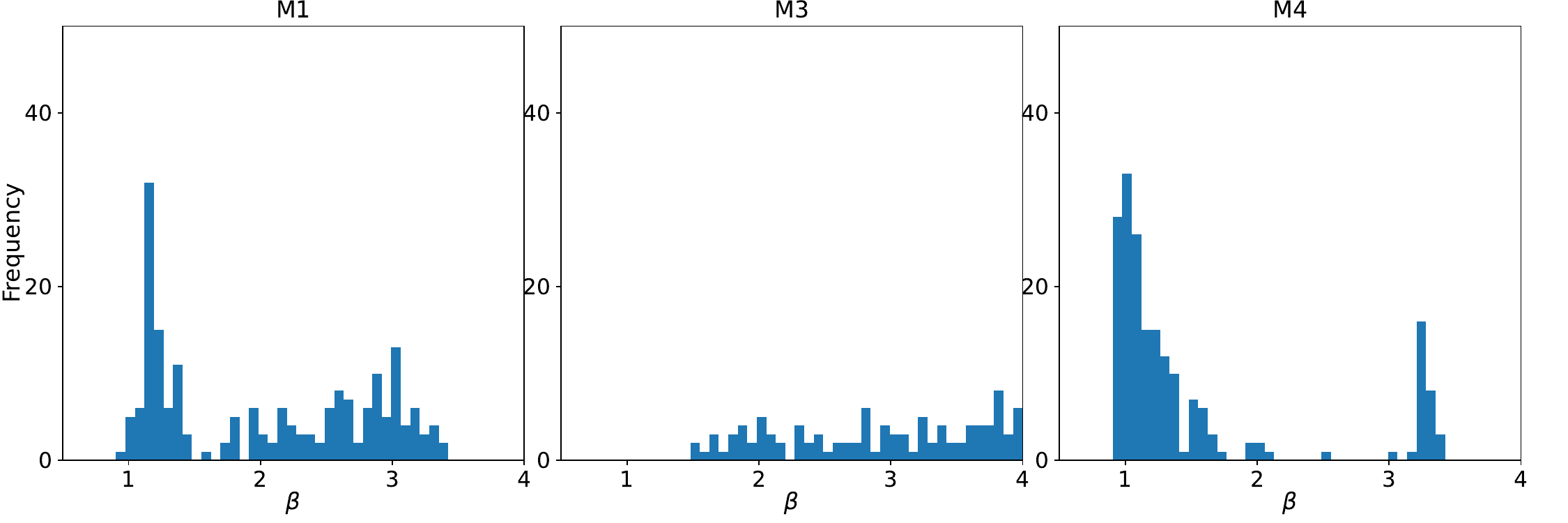}
    \label{fig:mse_single}
    } 
    \subfloat[MS-hyper GGM-c $\lambda$=0.0483]
    {  
        \includegraphics[width=0.49\linewidth]{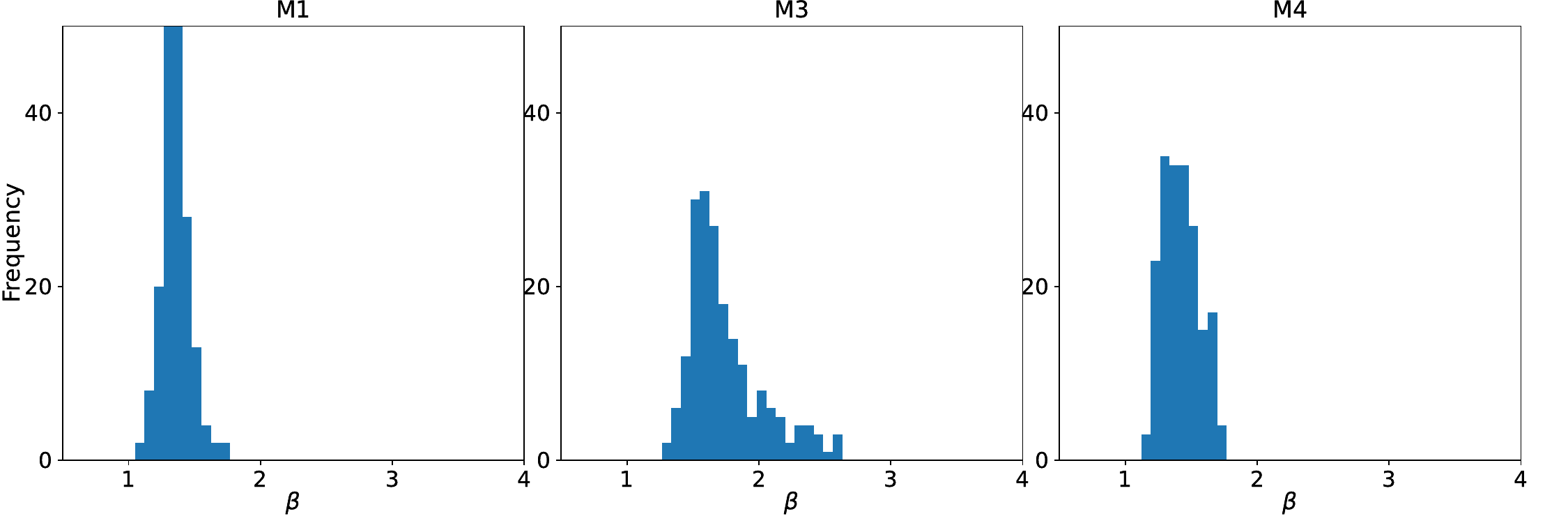}
        \label{fig:mse_single}
    }

    \subfloat[MS-hyper GGM-e $\lambda$=0.0018]
    {  
        \includegraphics[width=0.49\linewidth]{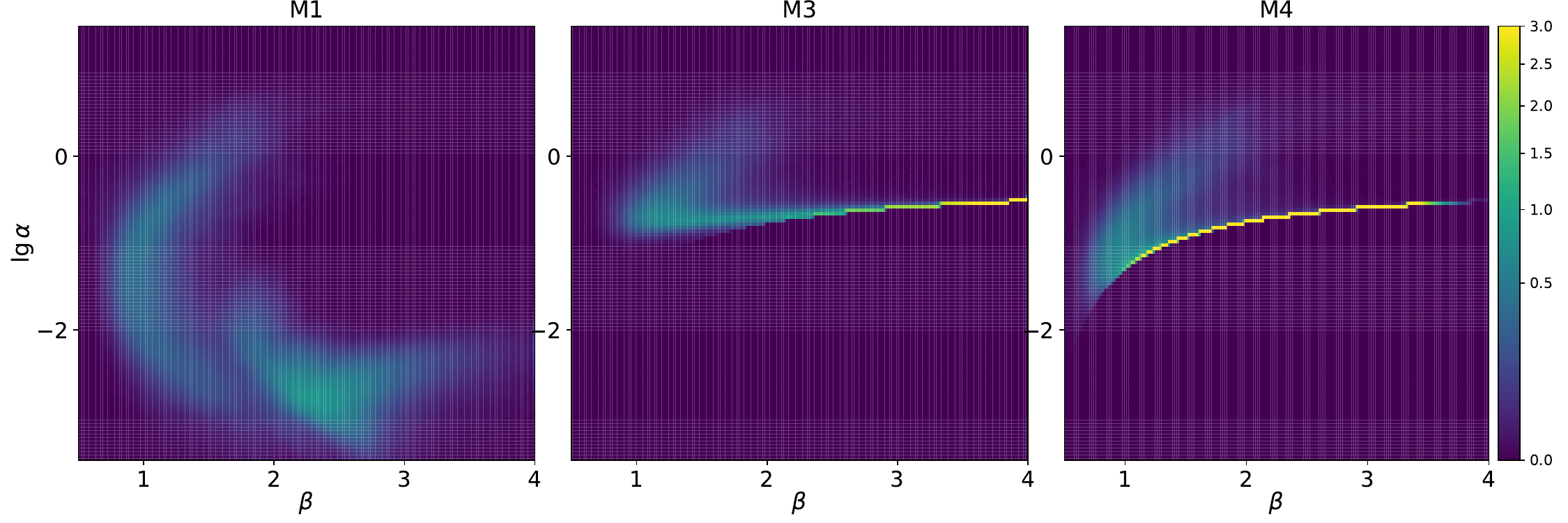}
        \label{fig:mse_single}
    } 
    \subfloat[MS-hyper GGM-e $\lambda$=0.0483]
    {  
        \includegraphics[width=0.49\linewidth]{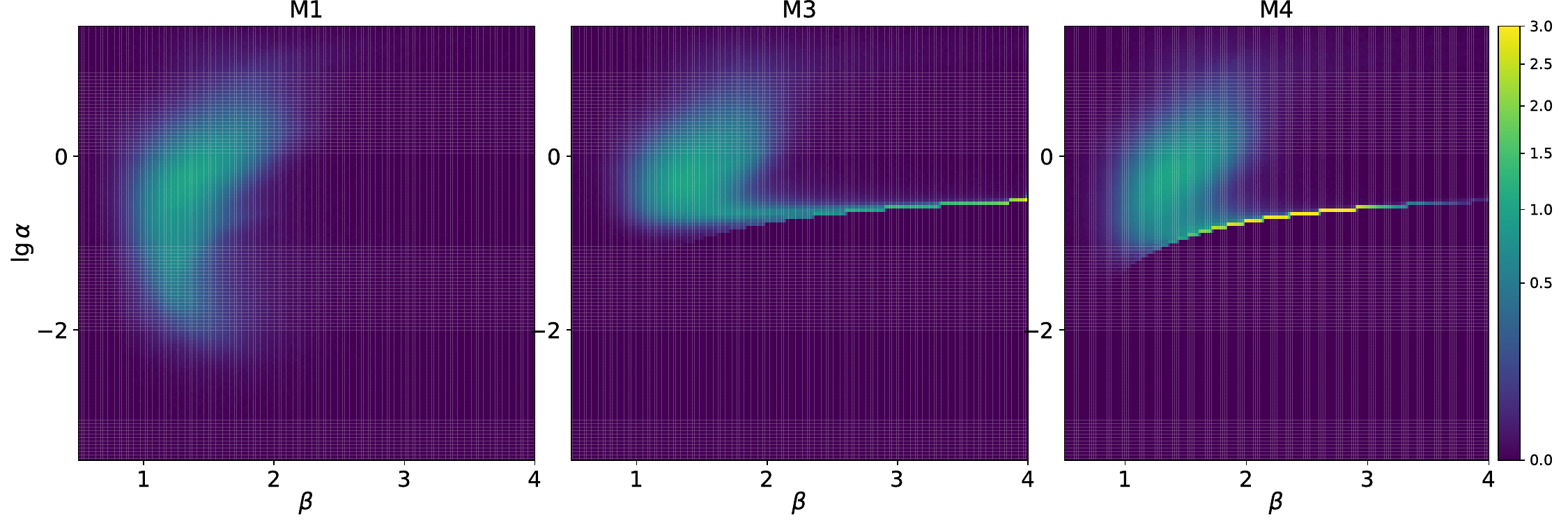}
        \label{fig:mse_single}
    } 
    
    \caption{
    Distribution of shape parameters $\beta$ and scale parameters $\alpha$ for ablation studies for improved training methods. (a) and (b) show the distribution of channel-wise shape parameters in MS-hyper GGM-c. (c) and (d) show the distribution of element-wise shape and scale parameters in MS-hyper GGM-e. $\lambda$=0.0018 represents lower bitrates ($<$0.2bpp) and $\lambda$=0.0483 represents higher bitrates ($>$0.8bpp). For MS-hyper, there are 192 channels. The results of GGM-e are collected from 7077888 latent variable samples in the Kodak dataset. The notations (M1, M3, M4) refer to Table \ref{tab:ablation}.
    }
    \label{fig:ablation_visual}

\end{figure*}

\subsubsection{$\beta$-dependent lower bound}
As shown in Fig. \ref{fig:ablation}, $\beta$-dependent lower bound significantly enhances the performance, especially at a lower bitrate. We use the distribution of $\beta$ and $\alpha$ parameters to better illustrate this method's effectiveness.
As shown in Fig. \ref{fig:ablation_visual}c and \ref{fig:ablation_visual}d, the estimated scale parameters concentrated on some smaller values if training without proper scale bound (M1). This causes a substantial train-test mismatch, resulting in poor performance, as discussed in Sec. \ref{sec:mismatch}. After incorporating the $\beta$-dependent lower bound, the estimated scale parameters could concentrate on larger values (M3), thus reducing train-test mismatch and improving performance. The values of scale parameters are much smaller at lower bitrates ($\lambda$=0.0018) than at higher bitrates ($\lambda$=0.0483). Therefore, the train-test mismatch is more severe at lower bitrates, and the performance at low bitrates can be more effectively improved by $\beta$-dependent lower bound.
The drawback of this lower bound is that it makes the shape parameter $\beta$ much larger than actually needed (M3), as shown in Fig. \ref{fig:ablation_visual}. This phenomenon is consistent with our analyses in Sec. \ref{sec:grad}. 

\subsubsection{Gradient rectification}
As shown in Fig. \ref{fig:ablation}, the gradient rectification method significantly improves the performance for GGM-m and GGM-c while bringing slight improvement for GGM-e.
Gradient rectification could mitigate the influence of the wrong direction gradient caused by $\beta$-dependent lower bound. As shown in Fig. \ref{fig:ablation_visual}, the distribution of $\beta$ after using gradient rectification (M4) concentrates on smaller values compared to only using scale lower bound (M3). It is more consistent with the original $\beta$ distribution of latent variables (M1).
For GGM-m and GGM-c, many latent variables share one shape parameter. Consequently, the wrong direction gradient will affect the optimization of all latent variables with the same shape parameter. However, for GGM-e, each element has its shape parameter. Thus, the influence of the wrong direction gradient in GGM-e is more minor than that in GGM-m and GGM-c. Therefore, gradient rectification is much more effective for GGM-m and GGM-c.
Additionally, since there are more latent variables at lower bitrates whose distributions lie below the lower bound, the influence of the wrong direction gradient is accordingly greater. Therefore, gradient rectification is more effective at lower bitrates to correct the wrong direction optimization.

\subsection{Influence of the Number of CDF Tables in LUTs-based Entropy Coding}
\label{sec:luts}
\begin{figure}[!t]

    \centering
    \includegraphics[width=0.95\linewidth]{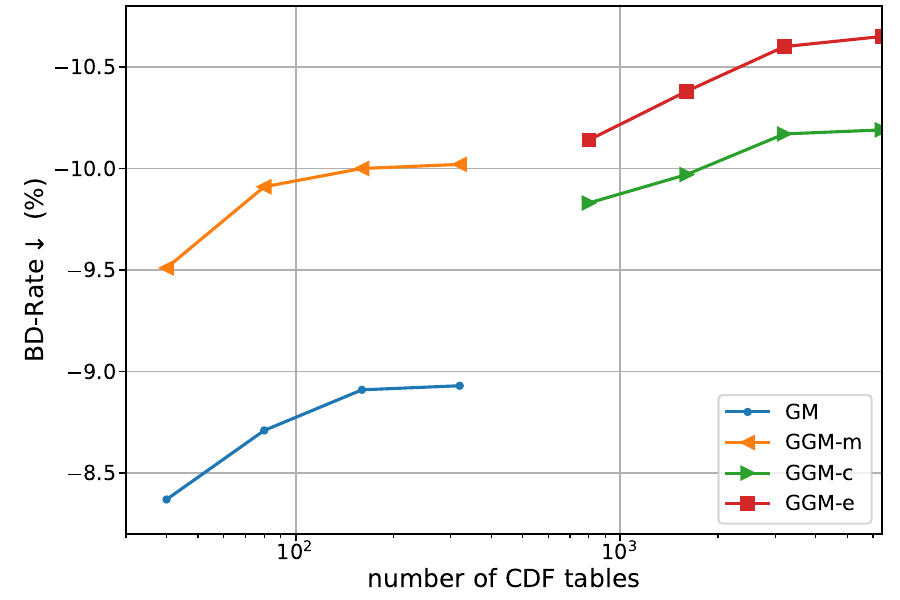}
    
    \caption{
    Performance of GM and GGM with different numbers of CDF tables in LUTs-based method. The experiments are conducted with the FM-intra model on the Kodak dataset. The anchor is the performance of VTM22.0.}
    \label{fig:lut}

\end{figure}
Figure \ref{fig:lut} shows the influence of the number of CDF tables on the compression performance. GM and GGM-m can achieve saturated performance with a few tables, while GGM-c and GGM-e require more LUTs to achieve saturated performance. We set the number of CDF tables in LUTs for GM and GGM-m as 160 and set 3200 for GGM-c and GGM-e due to the variety of shape parameters. The detailed results for determining the number of CDF tables are included in Appendix \ref{supp:5}. In summary, GGM-m achieves improved compression performance with the same number of LUTs as GM. GGM-c and GGM-e achieve further improved performance while requiring more LUTs.

\section{Conclusion}
In this work, we present a generalized Gaussian probabilistic model for learned image compression methods with more flexible distribution modeling ability and only one additional parameter compared to the mean-scale Gaussian model. With our further designed improved training methods for the generalized Gaussian model, which reduces the influence of train-test mismatch, we demonstrate the effectiveness of the generalized Gaussian model on a variety of learned image compression methods. Our GGM-m outperforms GM with the same network complexity and comparable coding time. Our GGM-c achieves better performance with the same network complexity and longer coding time. Our GGM-e achieves the best performance with higher network complexity and longer coding time. The increase in coding time for GGM-c and GGM-e is less than 8\% compared to that of GM. With the help of zero-center quantization and look-up tables-based entropy coding, our GGM-e outperforms GMM with lower complexity.

\bibliographystyle{IEEEtran}
\bibliography{main}

\appendices
\section{Derivation of the partial derivative of $c_{\beta}(y)$.}
\label{supp:1}
The incomplete gamma function $\gamma(r,z)$ and the regularized lower incomplete gamma function $P(r,z)$ are defined as 
\begin{align}
    &\gamma(r,z) =\int_{0}^{z}t^{r-1} e^{-t}dt,  \\
    &P(r,z) =\frac{\gamma(r,z)}{\Gamma(r)},
\end{align}
where $\Gamma(r)$ is the Gamma function.
Given the probability density function (PDF) of generalized Gaussian distribution, 
\begin{equation}
\begin{aligned}
     &f_{\beta}(y) = \frac{\beta}{2\Gamma(1/\beta)}e^{-|y|^{\beta}},\\
\end{aligned}
\end{equation}
the cumulative distribution function (CDF) can be derived through
\begin{align}
     c_{\beta}(y) = &\int_{-\infty}^{y}f_{\beta}(v) dv\\
     = &\int_{-\infty}^{y}\frac{\beta}{2\Gamma(1/\beta)}e^{-|v|^{\beta}} dv,\\
     &\text{define }w = |v|, \\
     = &\frac{1}{2}+\text{sgn}(y)\frac{\beta}{2\Gamma(1/\beta)}\int_{0}^{|y|}e^{-w^{\beta}} dw,\\
     &\text{define }t = w^{\beta}\\
     =& \frac{1}{2}+\text{sgn}(y)\frac{1}{2\Gamma(1/\beta)}\int_{0}^{|y|^{\beta}}t^{{\frac{1}{\beta}}-1}e^{-t} dt,\\
     =& \frac{1}{2}+\text{sgn}(y)\frac{1}{2\Gamma(1/\beta)}\gamma(\frac{1}{\beta},|y|^{\beta}),\\
     =&\frac{1}{2}+\frac{\text{sgn}(y)}{2}P(\frac{1}{\beta},|y|^{\beta}),
\end{align}
The partial derivative of $c_{\beta}(y)$ to $y$ can be derived through
\begin{align}
     \frac{\partial c_{\beta}(y)}{\partial y} = &\frac{\partial \int_{-\infty}^{y}f_{\beta}(v) dv}{\partial y}\\
     =& f_{\beta}(y)
\end{align}
Before calculating the derivative to $\beta$, we define 
\begin{align}
    r = \frac{1}{\beta},~z = |y|^{\beta}.
\end{align}
Then, the CDF can be written as 
\begin{align}
     c_{\beta}(y) 
     =& \frac{1}{2}+\text{sgn}(y)\frac{1}{2\Gamma(r)}\int_{0}^{z}t^{r-1}e^{-t} dt,\\
     =&\frac{1}{2}+\frac{\text{sgn}(y)}{2}P(r,z),
\end{align}
The partial derivative to $\beta$ is
\begin{align}\label{eq:chain}
     \frac{\partial c_{\beta}(y)}{\partial \beta} = &\frac{\partial c_{\beta}(y)}{\partial r}\frac{\partial r}{\partial \beta} + \frac{\partial c_{\beta}(y)}{\partial z}\frac{\partial z}{\partial \beta}.
\end{align}
Each component in Eq. (\ref{eq:chain}) can be calculated through
\begin{align}
    &\frac{\partial c_{\beta}(y)}{\partial r} = \frac{\text{sgn}(y)}{2} \frac{\partial P(r,z)}{\partial r},\\
    &\frac{\partial r}{\partial \beta} = -\frac{1}{\beta^2},\\
    &\frac{\partial c_{\beta}(y)}{\partial z} = \frac{\text{sgn}(y)}{2\Gamma(r)} z^{r-1}e^{-z},\\
    &\frac{\partial z}{\partial \beta} = |y|^{\beta}\ln{|y|}=z\ln{|y|}.
\end{align}
Then we have
\begin{align}
    \frac{\partial c_{\beta}(y)}{\partial \beta} = &\frac{\text{sgn}(y)}{2}\left(-\frac{1}{\beta^2}\frac{\partial P(r,z)}{\partial r}+\frac{z^r \ln{|y|}}{\Gamma(r)}e^{-z}\right)\\
    =&\frac{\text{sgn}(y)}{2}\left(-\frac{1}{\beta^2}\frac{\partial P(\frac{1}{\beta},|y|^{\beta})}{\partial 1/\beta}+\frac{|y|\ln|y|}{\Gamma(1/\beta)}e^{-|u|^{\beta}}\right)
\end{align}
For $\frac{\partial P(\frac{1}{\beta},|y|^{\beta})}{\partial 1/\beta}$, which does not have an explicit expression, we follow the implementation in TensorFlow.

\begin{figure*}[!t]
\centering
\subfloat[$X_1$]
{  
    \includegraphics[width=0.32\linewidth]{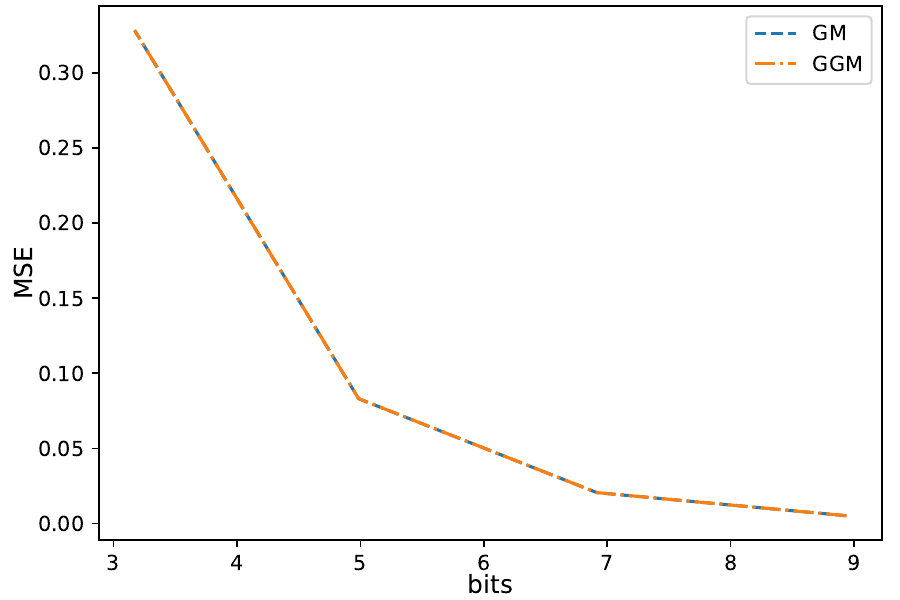}
}
\subfloat[$X_2$]
{  
    \includegraphics[width=0.32\linewidth]{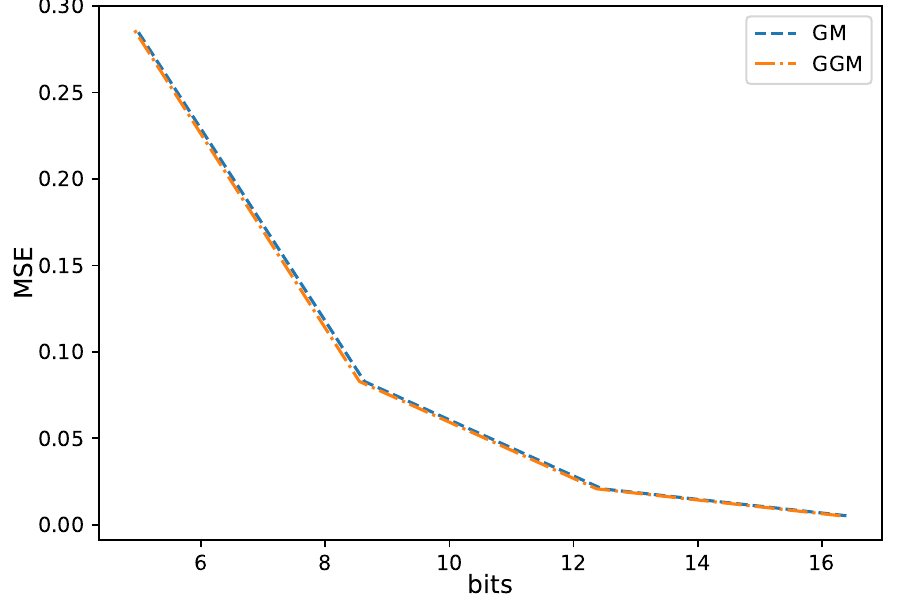}
}
\subfloat[$X_3$]
{  
    \includegraphics[width=0.32\linewidth]{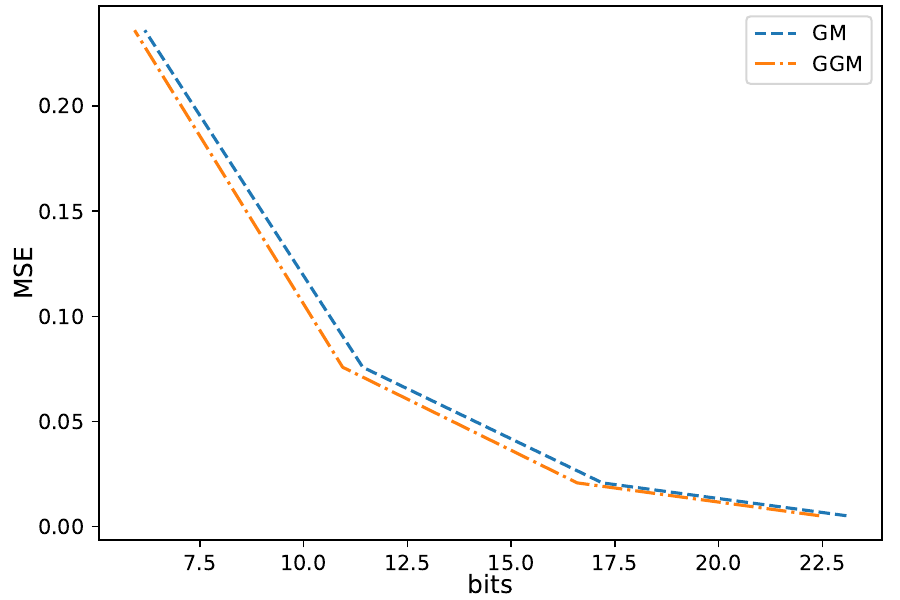}
} 
\caption{
Rate-distortion performance of GM and GGM applied to various sources.
}
\label{fig:theo}
\end{figure*}

\section{Simulation results with toy sources.}
\label{supp:2}
We have attempted to demonstrate the effectiveness of GGM through simple simulation examples. Specifically, we use two-dimensional stochastic sources with different distributions to illustrate the effectiveness of GGM within the transform coding framework. We use the optimal linear transform KLT of the corresponding source as the transform module followed by scalar quantization.  

We consider three types of source distribution as follows
\begin{equation}
\begin{aligned}
&X_1\sim p_{X_1}(x_1)=\mathcal{N}(0,\Sigma);\\
&X_2\sim p_{X_2}(x_2)=\frac{1}{2}\mathcal{N}(0,\Sigma)+\frac{1}{2}\mathcal{N}(0,\frac{1}{4}\Sigma);\\
&X_3\sim p_{X_3}(x_3)=\frac{1}{3}\mathcal{N}(0,\Sigma)+\frac{1}{3}\mathcal{N}(0,\frac{1}{4}\Sigma)+\frac{1}{3}\mathcal{N}(0,\frac{1}{16}\Sigma);\\
& \mbox{where }\Sigma=
\left[
\begin{array}{cc}
 2^2 & 1 \\
 1 & 1 
\end{array}
\right].
\end{aligned}
\end{equation}
For $X_1$, which is a Gaussian distribution, the coefficients after the transform still follow the Gaussian distribution, while for more complex sources $X_2$ and $X_3$, the transformed coefficients with KLT are no longer a Gaussian. For $X_2$ and $X_3$, GGM could more accurately model the distribution of the transformed coefficients. The rate-distortion performance when applying GM and GGM respectively is shown in Fig. \ref{fig:theo}.

For $X_1$, since the distribution of coefficients is Gaussian, the performance of GM and GGM is identical. For $X_2$, GGM achieves a 0.7\% rate saving compared to GM. For $X_3$, GGM achieves a 3.6\% rate saving compared to GM. The performance of GGM is highly dependent on the source distribution and the characteristics of the transform. We leave the exploration of the R-D bound achieved by GGM of more complex sources and transforms for future study.

\section{Fitting of $\beta$-dependent lower bound for scale parameter.}
\label{supp:3}
\begin{figure}[!t]
  \centering
  \subfloat[]
    {  
        \includegraphics[width=0.5\linewidth]{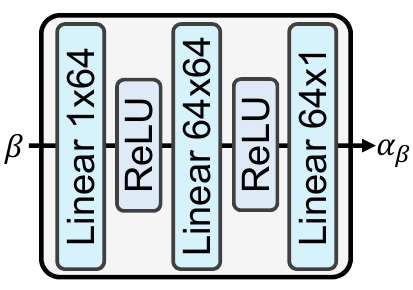}
        \label{fig:fc_scale_bound}
    } 
    \hspace{-1em}
    \subfloat[]
    {  
        \includegraphics[width=0.46\linewidth]{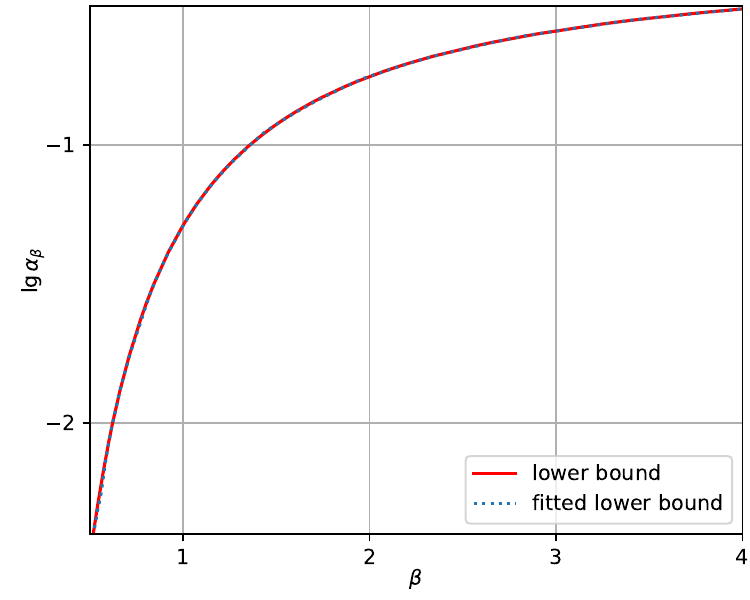}
        \label{fig:fit_result}
    } 
    \caption{
   (a) Network structure used to fit $\beta$-dependent lower bound for scale parameter. (b) Fitting result.
    }
    \label{fig:mse}
\end{figure}
The network structure used to fit the $\beta$-dependent lower bound is shown in Fig. \ref{fig:fc_scale_bound}. The fitted curve is shown in Fig. \ref{fig:fit_result}. We trained the network using the L1 loss function. We used the Adam optimizer with a learning rate of 1e-4. The network was optimized for 3000 iterations.

\section{Training cost of gradient rectification method.}
\label{supp:4}
A larger shape parameter can lead to numerical overflow during forward or backward propagation (gradient explosion). Our proposed gradient rectification method for GGM effectively eliminates incorrect gradients that would otherwise cause the shape parameter to grow unnecessarily large. Employing gradient rectification could result in more accurate shape parameters, indicating more precise distribution modeling, thereby improving compression performance.

\begin{figure}[!t]
  \centering
    \includegraphics[width=0.8\linewidth]{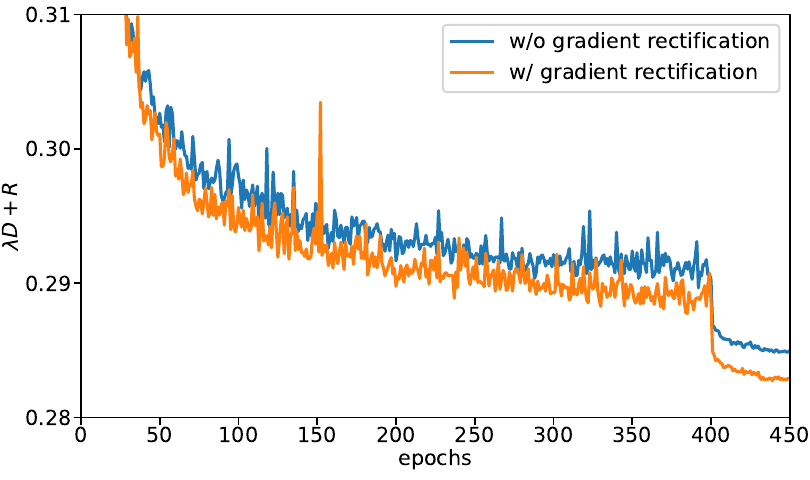}
    \caption{Loss convergence curves of training ELIC. ($\lambda=0.0018$).
    }
    \label{fig:loss}
\end{figure}

The average training time per iteration is shown in Table \ref{tab:gradient_rectify_elic}, which demonstrates that the gradient rectification method has little influence on the training time. The loss convergence curves are shown in Fig. \ref{fig:loss}. With the gradient rectification method, the rate-distortion cost converges to a better solution.
\begin{table}[!t]
\centering
\caption{Training time for per iteration of ELIC.}
\begin{threeparttable}
\begin{tabular}{cc} 
\hline
\Centerstack[c]{Method} & \Centerstack[c]{Time (ms)}\\
\hline
w/o Gradient rectification& 245  \\
w/ Gradient rectification& 247 \\
\hline
\end{tabular}
\end{threeparttable}
\label{tab:gradient_rectify_elic}
\end{table}

\section{Implementation details of look-up tables-based method for entropy coding.}
\label{supp:5}
First, we introduce the overview of the LUTs-based method for entropy coding. The LUTs-based method only influences the test stage and does not influence the training process. After training, several entropy parameter values of the probabilistic model are sampled in a specific manner from their respective possible ranges. We then calculate the corresponding cumulative distribution function (CDF) table for each sampled value. The encoder and decoder share these pre-computed CDF tables. During the actual encoding and decoding process, the predicted entropy parameters, which model the distribution of latent variables, are quantized to the nearest sampled value. These quantized values are then used to index the corresponding CDF table, which is necessary for entropy coders. For both GM and GGM, which have an identifiable mean parameter, we follow previous studies \cite{minnen2020channel,he2022elic,liu2023learned} to apply zero-center quantization, where the symbols encoded into bitstreams are given by $\lfloor y - \mu \rceil$. For GM, the entropy parameter involved in entropy coding is $\sigma$, while for GGM, the entropy parameters involved in entropy coding are $\beta$ and $\alpha$.

\begin{table}[!t]
\centering
\caption{Performance of LUTs-based entropy coding for GM on FM-intra\cite{li2024fm}.}
\begin{threeparttable}
\begin{tabular}{ cccc} 
\hline
\Centerstack[c]{Number of CDF} & \Centerstack[c]{BD-rate(\%)} &\Centerstack[c]{$t_e$ (ms)} & \Centerstack[c]{$t_d$ (ms)}\\
\hline
10&2.51&55.6&62.8\\
20&-6.69&56.8&63.2\\
40&-8.37&57.5&63.7\\
80&-8.71&58.9&64.9\\
\rowcolor{gray!20}
160&-8.91&64.6&68.6\\
\hline
\end{tabular}
\begin{tablenotes}
\footnotesize
\item[1] The settings we used are highlighted in color.
\item[2] BD-rate is evaluated relative to VTM22.0 on the Kodak set.
\end{tablenotes} 
\end{threeparttable}
\label{tab:lut_gm}
\end{table}

\begin{table}[!t]
\centering
\caption{Performance of LUTs-based entropy coding for GGM-e on FM-intra \cite{li2024fm}.}
\begin{threeparttable}
\begin{tabular}{ cccccc} 
\hline
\Centerstack[c]{$\beta$} & \Centerstack[c]{$\alpha$}& \Centerstack[c]{Number\\of CDF}& \Centerstack[c]{BD-rate(\%)} &\Centerstack[c]{$t_e$ (ms)} & \Centerstack[c]{$t_d$ (ms)}\\
\hline
10&10&100&15.96 &57.6 &65.1 \\
10&20&200&-5.71 &57.8 &64.9 \\
10&40&400&-9.22 &59.2 &67.4 \\
10&80&800&-10.14 &62.1 &68.9 \\
10&160&1600&-10.36 &69.8 &75.9 \\
20&10&200&15.60 &56.5 &64.6 \\
20&20&400&-5.99 &57.7 &64.7 \\
20&40&800&-9.48 &59.2 &66.5 \\
20&80&1600&-10.38 &62.4 &69.3 \\
\rowcolor{gray!20}
20&160&3200&-10.60 &69.8 &77.0 \\
40&10&400&15.53 &56.3 &65.0 \\
40&20&800&-6.05 &57.4 &65.2 \\
40&40&1600&-9.53 &59.4 &66.2 \\
40&80&3200&-10.44 &63.2 &70.5 \\
40&160&6400&-10.65 &71.0 &77.5 \\
80&10&800&15.51 &57.3 &66.2 \\
80&20&1600&-6.07 &57.8 &65.4 \\
80&40&3200&-9.55 &59.9 &67.2 \\
80&80&6400&-10.45 &64.0 &71.2 \\
80&160&12800&-10.66 &74.4 &80.7 \\
160&10&1600&15.51 &57.0 &66.1 \\
160&20&3200&-6.07 &58.2 &65.4 \\
160&40&6400&-9.55 &62.2 &68.8 \\
160&80&12800&-10.45 &69.5 &75.8 \\
160&160&25600&-10.66 &91.8 &95.3 \\
\hline
\end{tabular}
\begin{tablenotes}
\footnotesize
\item[1] The settings we used are highlighted in color.
\item[2] BD-rate is evaluated relative to VTM22.0 on the Kodak set.
\end{tablenotes} 
\end{threeparttable}
\label{tab:lut_ggm}
\end{table}

Then, we introduce the sampling strategy and entropy coding process for GM and GGM.
For GM, we linearly sample $M$ values in the log-scale field of $[0.11,60]$ for the $\sigma$ parameter. The sampling strategy is
\begin{align}
    \sigma_i = \text{exp}\left(\text{log}(0.11)+i\times\frac{\text{log}(60)-\text{log}(0.11)}{M-1}\right),
\end{align}
where $i\in\{0,1,\cdots,M-1\}$. We then calculate the corresponding CDF table for each $\sigma_i$. The precision of the CDF table is 16-bit unsigned integer (uint16) with a maximum length of 256 per table. Table \ref{tab:lut_gm} shows the performance of LUTs-based implementation for GM with different numbers of samples. To achieve better performance, we set $M=160$ for GM.
The total storage cost for these CDF tables is 0.08 MB. These 160 CDF tables are stored on both the encoder and decoder sides for entropy coding. During the test stage, the $\sigma$ parameter predicted by the entropy models is quantized to the nearest sampled value, which is then used to index the corresponding CDF table for entropy coding.

For GGM, several values for $\alpha$ and $\beta$ are sampled from their respective ranges. For GGM-m, since there is only one $\beta$ value, we only sample one $\beta$ value. In addition, we linearly sample 160 values in the log-scale field of $[0.01,60]$ for the $\alpha$ parameter. Since the range of $\alpha$ in GGM differs from the range of $\sigma$ in GM, the sampling interval for $\alpha$ is different from that of $\sigma$.

For GGM-c and GGM-e, the $\beta$ parameter is linearly sampled from the range $[0.5, 3]$ with $N$ samples, and the $\alpha$ parameter is linearly sampled in the log-scale range $[0.01, 60]$ with $M$ samples. Specifically, the sample values of the $\alpha$ parameter are 
\begin{align}
\label{eq:ggm_sample}
    \alpha_i = \text{exp}\left(\text{log}(0.01)+i\times\frac{\text{log}(60)-\text{log}(0.01)}{M-1}\right),
\end{align}
where $i\in\{0,1,\cdots,M-1\}$. We then combine the $N$ $\beta$ values and $M$ $\alpha$ values to generate $M \times N$ $\beta-\alpha$ pairs and calculate the corresponding CDF table for each pair. The precision of the CDF table is set to 16-bit unsigned integer (uint16), with a maximum length of each CDF table set to 256. These pre-computed CDF tables are stored on both the encoder and decoder sides for entropy coding. During the encoding and decoding process, the predicted $\beta$ and $\alpha$ parameters from the entropy models are quantized to the corresponding intervals respectively, which are then used to index the corresponding CDF tables.

Table \ref{tab:lut_ggm} presents the performance of the LUTs-based implementation for GGM-e with varying numbers of samples. For a balanced trade-off between complexity and performance, we set $N = 20$ and $M = 160$. For GGM-c and GGM-e, the storage cost for CDF tables is 1.56MB, which is relatively small compared to the network parameter counts. For instance, the network parameter size (using float32 precision) for the TCM model is 236MB, and the storage cost of the CDF tables constitutes only 0.6\% of the network size. The network parameter size for MS-hyper is 29MB, with the CDF table storage accounting for 5.3\%.

\begin{table*}[!t]
\centering
\caption{Computation time of cumulative distribution function of various probabilistic models using PyTorch.}
\begin{threeparttable}
\begin{tabular}{cccccccc} 
\hline
\multicolumn{8}{c}{Time (ms)}\\
\multirow{2}{*}{GM} & \multirow{2}{*}{GMM} & \multicolumn{6}{c}{GGM} \\
\cline{3-8}
&&\Centerstack[c]{$\beta=0.5$}&\Centerstack[c]{$\beta=0.75$}&\Centerstack[c]{$\beta=1.25$}&\Centerstack[c]{$\beta=1.75$}&\Centerstack[c]{$\beta=2.5$}&\Centerstack[c]{$\beta=3$}\\
\hline
16&38&22&22&22&22&21&21\\
\hline
\end{tabular}
\begin{tablenotes}
\footnotesize
\item[1] The computation time is 100 calculation processes of the CDF table with length as 256.
\end{tablenotes} 
\end{threeparttable}
\label{tab:CDF_time}
\end{table*}

\section{Entropy coding for GMM.}
\label{supp:6}
Since GMM has 9 parameters, using the LUTs-based implementation would incur excessive storage costs. For instance, sampling 20 values for each parameter in GMM would require storing $20^9$ CDF tables, which consumes at least $2\times10^5GB$ of storage. Therefore, we follow previous studies \cite{cheng2020learned,fu2023learned} to dynamically calculate CDF tables for GMM during encoding and decoding.

We implement the cumulative distribution function with PyTorch \cite{paszke2017automatic}. 
The comparison of computation time for CDF is shown in Table \ref{tab:CDF_time}. The results show that the CDF calculation time of GMM is longer than that of GM and GGM.

For the entropy coding of GMM, the CDF tables need to be dynamically generated for each latent element. Moreover, since the number of CDF tables for GMM equals the number of latent variables, which is significantly larger than that in the LUTs-based approach for GM and GGM, the time required for memory access becomes a notable factor. For instance, in the ELIC model, with an image resolution of $512 \times 768$, approximately 0.47 million CDF tables need to be calculated, resulting in a memory cost of 480MB when each CDF table has a length of 256. In contrast, GGM-c and GGM-e require only 1.56MB of memory. Thus, the memory access time for GMM is considerably higher. Due to the increased network complexity and the additional time costs associated with calculating CDF tables and memory access, the coding time for GMM is significantly longer than that for GM.

\section{Empirical results for determining the proper scale bound.}
\label{supp:7}
We follow the previous study \cite{zhang2023uniform} to empirically select the appropriate scale bound for other distributions. We have reported the experimental results for determining the optimal scale bound on MS-hyper, as shown in Table \ref{tab:bound}. The previous study \cite{zhang2023uniform} has also explored performance with different scale bounds across various learned image compression models. The scale bound for the Gaussian model, which serves as the primary comparison for GGM, has been widely adopted in the training implementations of recent advanced learned image compression models \cite{begaint2020compressai, liu2023learned, jiang2023mlic}.

\begin{table}[!t]
\setlength{\tabcolsep}{3.5pt}
\centering
\caption{Performance with different scale bounds on MS-hyper.}
\begin{threeparttable}
\begin{tabular}{ cc|cc|cc|cc} 
\hline
\multicolumn{2}{c|}{GM} & \multicolumn{2}{c|}{GMM} & \multicolumn{2}{c|}{LaM} & \multicolumn{2}{c}{LoM} \\
\Centerstack[c]{scale\\bound}&\Centerstack[c]{BD-\\rate(\%)}&\Centerstack[c]{scale\\bound}&\Centerstack[c]{BD-\\rate(\%)}&\Centerstack[c]{scale\\bound}&\Centerstack[c]{BD-\\rate(\%)}&\Centerstack[c]{scale\\bound}&\Centerstack[c]{BD-\\rate(\%)}\\
\hline
0.05&0.94&0.05&-0.04&0.01&1.1&0.005&0.64\\
0.08&-0.98&0.08&-3.44&0.03&-0.56&0.02&-1.16\\
\rowcolor{gray!20}
0.11&-1.81&0.11&-3.71&0.06&-1.62&0.04&-2.26\\
0.15&-1.61&0.15&-3.55&0.09&-1.61&0.07&-2.15\\
0.19&-0.41&0.19&-2.19&0.13&1.39&0.11&-1.01\\
\hline
\end{tabular}
\begin{tablenotes}
\footnotesize
\item[1] LaM represents the Laplacian model and LoM represents the Logistic model. 
\item[2] BD-rate is calculated relative to BPG.
\end{tablenotes} 
\end{threeparttable}
\label{tab:bound}
\end{table}

\vfill
\end{document}